\documentclass[structabstract]{aa} 
\usepackage{graphicx}
\usepackage{txfonts}
\usepackage{subfigure}
\usepackage{amssymb}
\usepackage{natbib}
\usepackage{url}

\def\kms{\rm {km\,s$^{-1}$}}
\def\x{$\times$}
\def\etal{et~al.}
\def\cmsq{\rm {cm$^{-2}$}}
\def\cmcub{\rm {cm$^{-3}$}}

\def\h2o{H$_2$O}
\def\so2{SO$_2$}
\def\13co{$^{13}$CO}
\def\nh3{NH$_{3}$}
\def\hco+{HCO$^{+}$}
\def\water18{H$_2^{18}$O}

\begin{document}    
\titlerunning {The first spectral line surveys   searching for  signals  from the Dark Ages} 
\title{The first spectral line surveys  searching for  signals  from\\ the Dark Ages
}

\authorrunning{C.M.~Persson \etal}   
   \author{C.M.~Persson
          \inst{1},
R.~Maoli\inst{\,2,3},
P.~Encrenaz\inst{4},
\AA.~Hjalmarson\inst{1},
M.~Olberg\inst{1, 5}, 
G.~Rydbeck\inst{1},
M.~Signore\inst{4}\\
 U.~Frisk\inst{6},
Aa.~Sandqvist\inst{7},
and J.Y. Daniel\inst{3}
}

   \offprints{carina.persson@chalmers.se}
   \institute{Onsala Space Observatory (OSO), Chalmers University of Technology, SE-439 92 Onsala, Sweden\\
              \email{\url{carina.persson@chalmers.se}}
\and Department of Physics, University of Rome "La Sapienza", Pl. A: Moro 2, 00185, Roma, Italy
\and Institut d'Astrophysique de Paris, 95 bis boulevard Arago, 75014 Paris, France
	\and LERMA , Observatoire de Paris and U.P.M.C., 61, av de l'Observatoire, 75014 Paris, France  
 \and SRON, Landleven 12, NL-9747\,AD Groningen, The Netherlands
\and Swedish Space Corporation, PO Box 4207, SE-171 04 Solna, Sweden
\and Stockholm Observatory, AlbaNova University Center, SE-10691 Stockholm, Sweden
}
\date{Received October 15, 2009; accepted March 8, 2010}

 
  \abstract
   {}
{Our aim is to observationally investigate the cosmic Dark Ages  in order to constrain star and structure formation models,  as well as the chemical evolution in the early Universe.
}
{{  Spectral lines  from atoms and molecules in primordial perturbations at high redshifts
can give information about the conditions in the early universe before and during the formation of the first stars in addition to the epoch of reionisation.  The lines may arise from moving primordial  perturbations before the formation of the first stars (resonant scattering lines), or could be thermal absorption or emission lines at lower redshifts.} 
The difficulties in these searches are that  the source redshift and evolutionary state, as well as molecular species and transition are unknown, which implies that  an observed line can fall within a wide range of frequencies. The lines are also expected to be very weak.
Observations from space have the advantages of stability and the lack of atmospheric features which is 
important in such observations. 
 We have therefore, as
a first step in our searches,  used  the Odin\thanks{Odin is a Swedish-led satellite
   project
   funded jointly by the Swedish National Space Board (SNSB), the Canadian Space Agency (CSA), the National 
   Technology Agency
   of
   Finland (Tekes) and Centre National d'Etudes Spatiales (CNES). The Swedish Space Corporation was the prime contractor and
   also is
   responsible for the satellite operation.} satellite to perform two sets of  spectral line surveys towards several positions.
The first survey covered the band   547\,--\,578\,GHz towards two positions, and the second one covered  the bands
542.0\,--\,547.5\,GHz and 486.5\,--\,492.0\,GHz towards six positions selected
to test different sizes of the primordial clouds.    
Two deep searches centred at 543.250 and 543.100\,GHz with 1\,GHz bandwidth were also performed towards one   position.
The two lowest rotational transitions of H$_2$ will be redshifted to these frequencies  from  $z\!\sim\!20\!-\!30$,    which is 
the predicted epoch of the first star formation. 
}  
{No lines are detected at an rms level of 14\,--\,90 and  5\,--\,35\,mK for the two surveys, respectively, and 2\,--\,7\,mK in the deep searches with a channel spacing of 1\,--\,16\,MHz.
The broad bandwidth covered allows a wide  range of redshifts to be explored for a number of atomic and molecular species and transitions.
From the theoretical side, our sensitivity analysis show that the largest possible amplitudes of the resonant lines  are   about  1\,mK 
at frequencies $\lesssim$\,200\,GHz, and a few $\mu$K around
500\,--\,600\,GHz, assuming optically thick lines and no beam-dilution. 
However, if existing, thermal  absorption lines have the potential to be orders of magnitude 
stronger than the resonant lines. 
We make a simple estimation of the sizes and masses of the primordial perturbations at their turn-around epochs, which previously has been identified as the most favourable epoch for a detection.  
This work may be considered as an important pilot study for our forthcoming observations with the Herschel Space Observatory.
}
   {}

   \maketitle

\section {Introduction}

One of the most important topics in astronomy today concerns the formation of the first stars and   structure formation in the Universe.
The cosmic microwave background radiation observed by  NASAs Cosmic Background Explorer\footnote{\url{http://lambda.gsfc.nasa.gov/}} (COBE) satellite and  the Wilkinson Microwave Anisotropy Probe$^1$  (WMAP)  shows small density fluctuations at a redshift of $z\!\!\approx\!\!1100$ and a temperature of $\approx$3000\,K. After this epoch of recombination the Universe became neutral and entered   the  cosmic Dark Ages which did not end until the formation of the first stars    and quasars, which are believed to have reionised the Universe at $z\! \sim \! 11$ \citep{2009ApJS..180..330K, 2009ApJS..180..306D}.  
During the Dark Ages the transition from the small density fluctuations left over from the inflation to completely formed objects took place  \citep[e.g.][]{2008arXiv0804.2258L}.  The challenging question is how and by what means we can reveal this process.

The main method to obtain information about physical and chemical conditions in  star forming regions at low redshift  is by means of atomic and molecular line observations. 
Molecular observations  of high redshift objects have also  been performed, 
for example by 
 \citet{2003A&A...409L..47B} who found 
high excitation CO  in a Sloan Digital Sky Survey quasar at $z\!=\!6.4$. 
However,  
no observational evidence exists at all  
from    the cosmic Dark Ages at even higher redshifts before or during the earliest epoch of the first star formation at predicted
redshifts of $\sim$20\,--\,30  \citep[e.g.][]{2005SSRv..117..445G, 2008IAUS..255....3G}.

Model-dependent theoretical analysis and computer simulations have therefore been the main tools to follow   perturbations of different scales in the    primordial medium to predict when and how the first stars and structure in the Universe evolved. This depends on a number of parameters including the chemistry in the Early Universe and
the properties of dark matter.
The chemistry constitutes a very important part in the Early Universe since star formation requires a cooling mechanism mainly provided by  molecules. 
Dust grains are very important for molecular production and for instance 
almost all of the molecular hydrogen production occurs today on   the surface of dust grains. 
The lack of dust and heavy elements in the Early Universe has therefore   resulted in  very low molecular 
abundances of only a few species and therefore   difficulties to explain the formation of the first stars \citep[e.g.]
[]{2004ARA&A..42...79B, 2005SSRv..117..445G}.

The Standard Big Bang Nucleosynthesis (SBBN) model predicts the formation of H, D, He and Li a few minutes 
after the Big Bang \citep[e.g.][]{2007ARNPS..57..463S}. Their primordial abundances in this model depend only 
on the density of baryons which has been measured with   high precision, 
$\Omega_\mathrm{b}$\,=\,0.0456$\pm$0.0015   \citep{2009ApJS..180..330K}. 
In the expanding and cooling Universe,  
molecular synthesis   of for example H$_2$, H$_2^+$, HeH$^+$,  HD, HD$^+$, LiH, and LiH$^+$ could begin as soon as neutral atoms appeared, attaining maximum abundances around redshifts of 100\,--\,400  depending on species \citep[e.g.][]{2009A&A...503...47V}.
The molecular primordial abundances  depend  on many parameters and
many attempts have been made to follow the chemical evolution during the Dark Ages  with results that often differ by orders of magnitude \citep[e.g.][]{1984ApJ...280..465L, 1993A&A...267..337P, 1998A&A...335..403G, 2002P&SS...50.1197G, 2002JPhB...35R..57L, 2006RoyalSocietyofChemistryB, 2007NewAR..51..411P, 2008A&A...490..521S, 2008MNRAS.387.1589S, 2009EPJC...59..117S, 2009ApJ...699..383B}.

The behaviour of gas in galaxy formation is still an open problem, despite the
development of  simulations, more precise semi-analytical models and more complete observational 
data during the recent years. In particular, cooling mechanisms in the primordial medium, star formation and feedback processes are far
from being completely understood \citep{2005SSRv..116..625C, 2006RPPh...69.3101B, 2007arXiv0712.2865E, 2008ASPC..393..111O, 2008ASPC..399..314D}.  In order to  be able to discriminate
between the different models of structure formation it is therefore   important to \emph{test theory with observations from the Dark 
Ages}. 

We do know that during this epoch, the Universe is filled with Cosmic Microwave Background (CMB)
photons, atoms, molecules, ions and electrons. The question is if these species have high enough abundances to
produce detectable signals. Since the average density on large scales is predicted to be very low,
$n_\mathrm{H}\!\sim\!0.2$\,\cmcub~at $z\!\sim\!100$  and decreasing towards lower redshift,
collisions are negligible and excitation is generally believed to be dominated by absorption of CMB
photons followed by spontaneous or stimulated emission in a process known as resonant scattering. This process 
alone is not able to produce a signal since CMB photons are nearly isotropic and resonant scattering
produce an isotropic distribution of photons. The only effect of resonant scattering by a primordial medium is the
damping of CMB primary anisotropies  \citep{1994ApJ...425..372M}. 
However, the gravitational potentials
induce 
peculiar velocities of the primordial clouds. Thus, the scattered CMB photons will be isotropically distributed in 
the reference frames of the clouds, but not in the CMB reference frame. As a consequence, secondary resonant
scattering anisotropies will be produced. The cross section for the resonant scattering is many orders of
magnitudes larger than for Thomson scattering between CMB photons and electrons, but it is extremely
frequency dependent, peaking at the rotational and rotational-vibrational frequencies of the scattering species.
Resonant lines will  therefore be produced by these moving primordial clouds \citep{1996ApJ...457....1M}.

Since the original idea by \citet{1977SvAL....3..128D},   inspired by earlier work on CMB fluctuations by Zeldovich,  a number of papers
have already analysed the possibility of detecting the resonant lines \citep[e.g.][]{1994ApJ...425..372M, 1996ApJ...457....1M,   2008NewA...13...28D}
and two previous observational attempts have been made without success \citep{1993A&A...269....1D, 2002ARep...46..543G}.

However, possible signals from the Dark Ages may not only arise from resonant scattering  since
the growing potentials may further induce adiabatic heating as well as shock heating of collapsing primordial perturbations  \citep[e.g.][]{2001PhR...349..125B, 2006ApJ...643...26S, 2006MNRAS.366..247J, 2008MNRAS.387.1021G}. 
This produces thermal emission or absorption from denser than average or high temperature regions.   
The  first stars are predicted to be very massive $\gtrsim$100\,M$_\odot$~
and formed either in isolation or at most as a small stellar multiple in minihalos \citep[e.g.][]{2000ApJ...540...39A, 2004ARA&A..42...79B, 2009Natur.459...49B}. These hot stars 
must have emitted enormous amounts of energetic radiation. 
By the time of reionisation, large 
HII regions are predicted to have grown around clusters of these hot and bright stars  \citep[e.g.][]{2008AIPC..990..405G}. This radiation will first destroy the molecules, but   in the cooling ejecta of the primordial supernovae    molecules will later be  efficiently produced  \citep{2008ApJ...683L.123C} for instance H$_2$   due to the high degree of ionisation. 
However, many considerations have to be taken into account to predict all these types of emission/absorption, hence  leading to  uncertain results.

We emphasize that every possibility to obtain  observations from the Dark Ages    should be seriously considered. 
A detection would be extremely important and would give \mbox{\emph{direct}} evidence of a very high redshift protostructure within a wide range of angular sizes, from arcsecond  to arcminute scales, depending on the cosmological model. 
These small scales are not observable in the CMB anisotropies due to the finite depth of the last scattering surface.
The horizon at the last scattering surface corresponds to 240\,Mpc today. The Planck\footnote{\url{http://planck.esa.int}} 
satellite, launched on May 14, 2009,   will measure the CMB power spectrum for angular scales greater than 10\arcmin~(40\,Mpc). Smaller scales, at cluster and galaxy sizes are unobservable due to photon diffusion damping associated with the non-instantaneous nature of the recombination process at the last scattering surface \citep{1968ApJ...151..459S}.

Any primordial spectral line will also provide a unique test of nuclear synthesis at high redshift, give new information about the reionisation of the universe, probe the chemistry throughout a wide range of redshifts, the heating and cooling processes as well as the dynamics of the primordial clouds before and during the gravitational collapse of a protostructure.     
Even a non-detection could  give very valuable information in order to constrain all  these   issues if the noise level is low.

Given the expected weakness of the lines and our ignorance of the frequencies at which the transitions will fall, 
ground-based observations are complicated by  the terrestrial atmospheric lines. The ozone molecule, O$_3$, including all the isotopes $^{16}$O, $^{17}$O and $^{18}$O,  emits numerous lines. Moreover, most of the vibrationally excited lines   can  not be found in  catalogues.
Observations with the IRAM 30-m telescope  \citep{1993A&A...269....1D} experienced   difficulties from these lines at a low level.
Interferometer observations are therefore interesting for searches like this, due to the
 many
advantages for rejecting both broad   and narrow band atmospheric
emissions.  Ground based single dish observations can, however, also provide valuable information  especially at low frequencies, even though the foreground
and atmospheric radiation poses a large problem. 
Another possibility is
observations from space, which have the advantages of stability and  lack of atmospheric features. 

As a first test, we have therefore chosen   to  use a satellite to search  for and put upper limits on   primordial  signals from the Dark Ages.   
The satellite is  required to have tunable receivers in order to cover a broad spectral band,  
essential for the exploration of
a wide redshift range.
In this work, we have  used the only available satellite at the time with the above mentioned requirements: the Odin satellite \citep{2003A&A.402.21.Nordh.etal}. This satellite  has the unique capability to cover  a broad spectral band  with a high channel spacing of about 1\,MHz
using  tunable single sideband  (SSB) receivers. This has allowed us for the first time to search for redshifted molecular hydrogen from  the predicted time of the formation of the first stars or earlier.


\section{The Odin observations}

The main problems in the searches of primordial lines  are the unknown but   expected  weak  amplitudes from unknown species in unknown sources    at unknown   redshifts.  This implies that 
the frequency of the lines can be anywhere in a wide frequency range.  Thus we need to cover a large frequency band to probe a wide range of redshifts for many possible species and transitions. The sources should on the other hand be found in every direction on the sky even though they most likely have a clumpy distribution and also are located at different redshifts. This implies that  we do not know the evolutionary state, size or density of the perturbations.

\subsection{Interesting species}

Our approach in all  observations  has been   to  perform  spectral line surveys
in order to cover a wide redshift range. This is  necessary since
our objects are located at an unknown redshift and
the expansion of the Universe causes the frequency of an emitted photon to be redshifted as
\begin{equation}\label{redshifted freq}
\frac{\nu_\mathrm{0}}{\nu_\mathrm{obs}} = 1+z \ ,
\end{equation}
where ${\nu_0}$ is the rest frequency, $\nu_\mathrm{obs}$ the observed frequency of the transition, and $z$ is the redshift. 
When we choose an observation frequency, we thus determine the emitting redshift for each transition given  by Eq.~(\ref{redshifted freq}). 
Each transition may also be detected at different frequencies,  emitted from several objects at different redshifts along the line of sight. 
The strong frequency dependence of the scattering and emission processes gives rise to lines superposed with the black body continuum spectrum of the CMB. Each line is associated with a moving primordial perturbation, exactly in the same way the  \mbox{Lyman $\alpha$} absorption lines are associated with neutral hydrogen clouds which absorb  the light from bright quasars and galaxies.

We search for any   atom or molecule  that may be present in the high redshift Universe. This includes
molecules that   form  from the primordial elements H, D, He and Li \citep{2007ARNPS..57..463S} such as: H$_2$, H$_2^+$, H$_2$D$^+$, HD, HD$^+$, HeH$^+$,  
LiH and LiH$^+$.
There are also  suggestions that lines from neutral hydrogen H and  helium He could be seen during the time of respective recombination and at later times  \citep[][]{2006MNRAS.371.1939R, 2006A&A...458L..29C, 2007ApJ...664....1S, 2007MNRAS.374.1310C, 2008A&A...485..377R}.
Prior to the reionisation at $z \! \gtrsim\! 6$ the universe was, however, opaque to for instance Ly$\alpha$ radiation\citep{2001AJ....122.2850B}.

Previous searches for primordial  resonant  lines were mainly focused on LiH due to its  high dipole moment and a possibly high abundance  \citep[][]{1990ApJ...357....8D}. New ab initio calculations of reaction rates of LiH destruction in the early universe show, however, that the LiH   abundance is too low to allow  a detection \citep{2009ApJ...699..383B}. Another interesting molecule with a high dipole moment is HeH$^+$ which, together with a much higher abundance than LiH, increases the possibility of a detection.

Moreover,  non-standard BBN models   predict  heavy nuclei such as carbon, 
oxygen,   nitrogen, and flourine, which may have created   molecules like CH, OH, NH,  HF, and their respective 
molecular ions \citep{2007A&A...476..685P, 2007IAUS..235..413C, 2008sf2a.conf..355V, 2009A&A...503...47V}.
For example the abundance of the CH molecule is
found to be higher than the H$_2$D$⁺$ or HD$⁺$ abundances in the  inhomogeneous non-standard BBN models.
In effect, a number of possible mechanisms have been suggested to generate density perturbations in the early 
universe which could survive until the onset of primordial nucleosynthesis.  Such 
inhomogeneities in the baryon number may result from non-equilibrium processes in the Big Bang, for example, 
occurring
during a putative first-order quantum-chronodynamics (QCD) phase transition at $\sim$100\,MeV or during the 
electroweak symmetry breaking at $\sim$100\,GeV. Assuming the existence of such surviving density 
fluctuations, some authors have shown that -- in the framework of these non-standard BBN scenarios -- the 
altered primordial nucleosynthesis could lead to the synthesis of heavy elements
\citep[e.g.][]{1987PhRvD..35.1151A, 1994ApJ...429..499R, 1997ApJ...479...31K,2006PhRvD..73h3501L}. 
Note also that these models   satisfy the observed primordial elemental light
abundances.

In addition, regardless of BBN model,   the “first generation” stars of a primordial composition  formed at    $z\!\ga\!20\!-\!30$ might quickly have polluted the  medium with small amounts of metals with subsequent chemistry \citep[e.g.][]{2003ApJ...586....1M, 2004ARA&A..42...79B, 2009ApJ...703..642C, 2009ApJ...691..441S}. Several species have already been observed at high redshifts.
In addition to the high excitation CO observations at $z\!=\!6.4$ 
\citep{2003A&A...409L..47B}   this molecule has also   been observed 
in for example a  damped Lyman-{$\alpha$} system at $z$\,=\,2.4 \citep{2008A&A...482L..39S} and  in 
a field containing an over-density of Lyman break galaxies at z\,=\,5.1 \citep{2008ApJ...687L...1S}. 
A massive CO reservoir has also been detected   at $z$\,=\,3.9  \citep{2001Natur.409...58P}.  
An unusually high amount of neutral hydrogen was found by \citet{2008ApJ...685L...5F} in a 14\,Mpc region surrounding a young galaxy at z\,=\,4.9. A strong detection of molecular absorption bands, including H$_2$ and CO, was also recently observed from gas within the host galaxy of the gamma-ray burst 080607 at $z$\,=\,3.0363 \citep{2009ApJ...691L..27P}. Several absorption features are still unidentified, but a H$_2$ column density of 10$^{21.2\,\pm\,0.2}$\,\cmsq~was inferred with an excitation temperature of 10\,--\,300\,K. The highest redshift object observed today is the gamma-ray burst of 23 April 2009 at $z$\,=\,8.2 \citep{2009Natur.461.1254T}. 
The long duration gamma-ray bursts are believed
to be primordial supernovaes originating from the super-massive first generation stars,  i.e. population III stars \citep{2006ARA&A..44..507W}.

The continuum of H$^-$ has also been suggested to give rise to a detectable signal in terms of a decrease in the CMB spectrum which may superimpose a detectable absorption feature on the CMB. This could be detectable with the Planck satellite, although the 
strength of this effect has  been estimated to quite different values \citep{2006RoyalSocietyofChemistryB, 2008A&A...490..521S}.

Besides producing molecular lines, resonant scattering has an additional interesting feature   that is 
regardless of peculiar velocities,
this process   has the potential of reducing the power of the CMB primary anisotropies  if the molecular abundances are high enough \citep{1994ApJ...425..372M}. Planck  may be able to detect also this decrease in power. There are also several suggestions to use
differential measurements of the broad band CMB angular power spectrum observed with  Planck in the search for  resonant lines \citep[e.g.][]{2004A&A...416..447B, 2008NewA...13...28D, 2008A&A...490..521S}.  
To reach the required sensitivity of the order of $\mu$K, the main problem to extract these weak signals with this approach, will be the galactic and extra-galactic foreground emissions.
 
 Since the most abundant species is neutral hydrogen, a very interesting and promising transition is the 21\,cm line.
Such difficult observations must, however, await the future
Low-frequency 
Array\footnote{\url{http://www.lofar.org/}} (LOFAR) and the proposed Square Kilometer Array 
(SKA)\footnote{\url{http://www.skatelescope.org/}} which will be able to
detect the redshifted \mbox{21-cm} H{\sc I} transition between \mbox{$z\!\sim\!6\!-\!11$} or at even higher redshifts  \citep[e.g.][]{1979MNRAS.188..791H, 2009astro2010S..83F}. 
Both instruments, however,  face severe difficulties when subtracting the astrophysical foreground contamination which is  several orders of magnitudes stronger than the 21cm line \citep{2009ApJ...695..183B}.

Most other atomic transitions than the 21\,cm line have much higher required excitation temperatures than the molecular transitions and, except for H, He and Li, must in addition have been produced either by the first stars or in a non-standard BBN. They 
are therefore only expected to be seen
at redshifts during the epoch of first star formation or some time after. As a first step, we have therefore concentrated our searches on optically thin primordial molecular lines. The redshifted frequencies from  $z\!\sim$20\,--\,30 of the lowest transitions of \emph{ortho}- and \emph{para}-H$_2$  fall at Odin frequencies around 500\,GHz. These lines could  be observable  in absorption towards the CMB or more likely towards  for example hot HII regions. 
The detectability of H$_2$ rotational lines in emission  associated with the formation of the first stars has previously  been discussed   \citep{2003ApJ...599..738O, 2003MNRAS.339.1256K} in addition to the  ro-vibrational H$_2$ emission lines   \citep{2004PASJ...56..487M}.
In addition, a number of lines of for example  H$_2^+$, H$_2$D$^+$, HD, HD$^+$, and HeH$^+$ can also be seen from this epoch.
 
At redshifts higher than about $z\!\gtrsim\!150-200$ the kinetic matter temperature   is expected to be the same as the radiation temperature  of the CMB  due to Compton scattering,
evolving as  
\begin{equation}
T_\mathrm{CMB}= T_\mathrm{K} = T_0(1+z)\ ,
\end{equation}
where $T_0$\,=\,2.725\,K is
the    cosmic microwave background radiation (CMB) today, as measured   by the   FIRAS instrument on-board the COBE satellite \citep{1996ApJ...473..576F}. During this epoch only resonant scattering may occur since temperature differences are 
required
to produce thermal spectral lines  (see Eq.~\ref{solution}).
At  redshifts below 200, the matter temperature evolves following an  adiabatic expansion which
implies a kinetic temperature decrease faster than for the CMB  \citep{2005SSRv..117..445G}
\begin{equation}\label{TK decrease}
T_\mathrm{K} \propto    (1+z)^2. 
\end{equation}
These temperature differences will force the excitation further away from 
equilibrium with the radiation,  and should enhance spectral absorption 
distortions 
at redshifts 
\mbox{$\lesssim$\,150\,--\,200.}
The gas will eventually be heated by  the accretion and collapse phases of the growing perturbations, and \mbox{eventually} 
by the first stars and quasars which will reionise the whole universe at $z\!\! \gtrsim 6$. This will enhance spectral emission features.

The coupling of the excitation temperature to the radiation or matter temperature differs  with species.
Polar molecules like HeH$^+$ probably continue to have $T_\mathrm{ex}$ coupled to
the CMB temperature, while  H$_2$   may remain more strongly coupled to the matter
temperature. There are also situations where 
the excitation temperature may  depend on molecular formation processes, by which for instance primordial
H$_2$ may remain superthermal relative to  
the matter temperature.

The strength of a spectral line depends, in addition to the number density and dipole moment, on the ratio of the   upper state energy 
(in Kelvin) 
and excitation temperature (see Eq.~\ref{tau}). 
Rotational-vibrational lines, with upper state energies of thousands of Kelvin, are  therefore  only expected for very high redshift sources with a high temperature or from
  collapsing high-temperature regions at lower redshifts. In such regions the density may   be increased above average, and
the increased collision rate in these regions will drive
the level populations and the excitation temperature towards the high matter temperature.

A detection of several spectral lines is required to secure a molecular identification and a redshift of an emitting source.
The spectral distance between rotational lines is, however, rather wide, even though it becomes increasingly more narrow
for highly redshifted sources observed below 100\,--\,200\,GHz.
It is thus more likely that a single transition from each object  would be seen in the Odin observations.
The rotational-vibrational transitions would  be more closely spaced, but have a   lower probability of detection.

\subsection{Observational strategy}

Neither the amplitude nor the line widths are known and therefore 
we  use   high spectral resolution to enable  detection  of narrow lines, and  later re-bin the data to lower resolution.
The choice of spectral resolution  affects the root-mean-square noise level $T_\mathrm{rms}$
as can be
seen from the radiometer formula
\begin{equation}\label{radiometer formula}
\frac{T_\mathrm{rms}}{T_\mathrm{sys}} = \frac{K}{\sqrt{t\, \Delta f}} \ ,
\end{equation}
where $ \Delta f$ is the frequency resolution of the spectrometer, $t$ is the on-source integration time,  $T_{sys}$ is the noise temperature from the whole system, which includes the receiver noise, and $K$ is a constant which depends on the observation strategy ($K\!=\!\sqrt{2}$ for a switched receiver).

 The expected signals most likely originate from every direction in the
sky, even though they will be stronger from rare high density
peaks. However, we needed to observe positions with as low contamination
as possible from any known source, and in addition they had to be
observable with the Odin satellite for long periods. For these reasons
we have observed towards two of the WMAP hot spots in the CMB radiation,
out of the Galactic plane. The coordinates are listed in
Table~\ref{coordinates table}.

\begin{table} [!h]
\caption{Coordinates of observations.}
\begin{tabular} {   l l l }
  \noalign{\smallskip}
\hline
\hline
 \noalign{\smallskip}
Position & R.A. ($J$2000)  & DEC. ($J$2000)  \\
   \noalign{\smallskip}
\hline
Hot Spot 1 & 05$^h$26$^m$00$^s$0 & $-$48$\fdg$\,30$\farcm$\,00$\farcs$0\\
Hot Spot 2 & 05$^h$09$^m$36$^s$0 & $-$43$\fdg$\,24$\farcm$\,00$\farcs$0\\
   \noalign{\smallskip}
\hline
\end{tabular}
\label{coordinates table} 
\end{table}

Two different sets of observations were performed  with different observation strategies as described below. 
The first survey took place  in 2004 and the second one in 2006\,--\,2007. In addition, we performed two sets of deep searches at one frequency setting towards one position in 2007 and 2009.

The Odin \mbox{1.1 m} offset Gregorian telescope has a circular beam at 557\,GHz and a FWHM beam width of  2$\farcm$1 \citep{2003A&A.402.27.Frisk.etal}. This corresponds for example to 540 and 78\,kpc at $z$\,=\,10 and $z$\,=\,100, respectively (see also Table~\ref{Odin step scale}, on-line material).
Being outside the atmosphere, and with an exceptionally high
main beam efficiency, $\eta_{\mathrm{mb}}$\,=\,0.9, our  intensity  calibration is very accurate.  
The
calibration procedure (the chopper wheel method) of the Odin satellite is described in \citet{2003A&A.402.35O.Olberg.etal}. 
The intensity scale in the figures is expressed in terms of antenna temperature $T^*_\mathrm{A}$. 
The reconstructed pointing offset was $\lesssim$15$\arcsec$~during most of the time.
 
Two different single sideband (SSB) receivers, with a typical side-band suppression of $\gtrsim$20\,dB,  were used simultaneously in combination 
with one auto-correlator (AC) and one acousto-optical spectrometer (AOS).
The bandwidth   is 1.040\,GHz for the AOS, and 690\,MHz for the AC,
with a channel spacing of 0.62 and 1.0\,MHz for the AOS and AC, respectively, and thus 
$\Delta \nu / \nu\!\approx\!2\!\times\!10^{-6}$ at $\nu$\,=\,545\,GHz.
The Odin   average system temperature is around 3\,300\,K (SSB) and therefore many hours of observations are demanded to produce sensitive observations. The resulting noise levels may very  well be  too high to allow a detection,
but the intensity of the lines is really not known and we have therefore taken a pure observational approach. If no detection is obtained, the resulting upper limits will be used as input   to  the  noise levels 
required in  future observations. This approach also applies to the chosen frequency resolution which is connected to the noise level and final covered bandwidth.

In each Local Oscillator (LO) setting, 
the resulting AOS spectra have a very stable baseline except at the band edges   where the calibration spectra exhibit a steep increase.  The noisy edges are excluded, and 
then a first or second order polynomial baseline is subtracted before we join all spectra together to form a  contiguous spectrum. 
The AC consists of 7 bands of 100\,MHz each in each setting. We subtract a linear baseline in each band before we align and average all spectra.   All averages are then joined.

\subsubsection{Observations during  summer 2004}

Our first goal was to 
cover as wide a frequency band as possible. We were awarded 337 orbits (1 orbit$\sim$1 hour of observation) and performed independent position switching  observations towards the two positions listed in  Table~\ref{coordinates table},  
using
reference positions   offset by -45~arcminutes  in declination.
Using simultaneous observations of the AOS and AC,  15.5\,GHz were covered in steps with each receiver:  AC  \mbox{$\sim$563\,--\,578 GHz} and AOS \mbox{$\sim$547\,--\,563 GHz}, giving at total observed frequency band of 31\,GHz   covering the full \mbox{547\,--\,578\,GHz} band. We spent 5 orbits for each  LO setting and the step size in frequency was 0.5\,GHz.
The
settings and steps were the same as in the first 
spectral line survey performed by the Odin satellite towards the Orion KL nebulae \citep{2007A&A...476..791O, 2007A&A...476..807PaperII}.

\subsubsection{Observations during 2006/07}

\begin{figure}[\!ht] 
\centering
\includegraphics[scale=0.2]{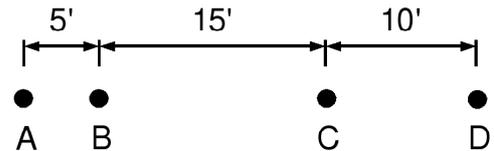}   
 \caption{Observation  strategy during the second spectral survey 2006/07 (also described in Table~\ref{Table with points and spacings}).  The Odin beam has a FWHM width of 2\farcm1. }
 \label{obs strategy 2006-2007}
\end{figure}

In the next attempt during nine weekends in winter 2006/07 –and one weekend  in  Aug 2007 (in total 464 orbits) we changed our observational strategy. We wanted to  lower the noise compared to what we obtained in the 2004 observations, but still needed to cover a wide frequency band. 
Since the sizes of the primordial objects are unknown, this time   
we also wanted to test different spacings between the signal  and reference positions in order to avoid subtraction of the signal if  present in both positions. 
A compromise of the above requirements resulted in the scheme
shown in Fig.\,\ref{obs strategy 2006-2007}. Four positions,
\mbox{A\,--\,D}, were observed where
position A is   the previously   \mbox{Hot Spot 1} observed  during 2004. The minimum offset was determined by the Odin beam size of 2$\farcm$1, and thus 
we chose the following offsets:  $-$5\arcmin, $-$15\arcmin, and $-$10\arcmin~in DEC, keeping the same R.A.
During 10 weekends of observations ($\sim$40 orbits/weekend) we observed all four positions with one frequency setting each weekend. Table~\ref{Table with points and spacings} describes how we performed the observations: 
10 orbits towards each pair of  positions were observed in a sequence.  First towards position A with  B as reference (AB),  then B with C as a reference (BC),    C with D as reference  (CD), and finally   D with A as a reference (DA). Thus, in total we have
ten orbits/setting/position for ten different LO settings.


\begin{table}[\!ht] 
\centering
\caption{Observation strategy  2006/07: every weekend we observed four positions, \mbox{A\,--\,D}, in a sequence using $\sim$10 hours towards each position with one frequency setting   (also shown in Fig.\,\ref{obs strategy 2006-2007}).
}
\begin{tabular} {ll ll } 
 \hline\hline
     \noalign{\smallskip}
 	&		Sig	&	Ref	&	Spacing	\\
     \noalign{\smallskip}
     \hline
     \noalign{\smallskip}
1	&	A	&	B	&	5\arcmin	\\
2		&	B	&	C	&	15\arcmin	\\
3		&	C	&	D	&	10\arcmin\\
4		&	D	&	A	&	30\arcmin	\\
     \noalign{\smallskip}
\hline 
\label{Table with points and spacings}
\end{tabular}
\end{table} 

 As in the first observation run in 2004, we performed simultaneous observations with the AC and the AOS, but 
covering other frequency ranges.  
The final results are spectral surveys in the frequency ranges \mbox{542.0\,--\,547.5\,GHz} (AOS) and \mbox{486.5\,--\,492.0\,GHz} (AC) towards four positions.

 \subsection{Deep searches at 543\,GHz during 2007 \& 2009}
 
In addition to the spectral surveys we also wanted to perform a deep search to lower the noise even further. 
In 2007 we therefore used 40 orbits towards HotSpot1 position A with B as reference position with 543.250\,GHz 
as centre frequency with the AOS and 490.250\,GHz with the AC. 
This produced an rms level of about 10\,mK over the 1\,GHz band with a channel spacing of 1\,MHz, and a 
possible 4\,$\sigma$ detection at 543.1\,GHz.
To validate the possible detection, 
we performed additional observations in April 2009 using 96 orbits and then changed the centre frequency of the 
AOS  to be 543.1\,GHz.

\section{Results}\label{section results}

No lines were  detected and   we are thus limited to set upper limits of possible signals. We choose a  5$\sigma$ 
noise level as an upper limit   since the peak noise level is always lower than this value if the noise follows a 
Gaussian distribution and has independent velocity channels. In this case, the probability that an observed 
5$\sigma$ line would be a noise feature is less than 10$^{-6}$.

\begin{figure}[\!ht] 
\centering
\resizebox{\hsize}{!}{\includegraphics{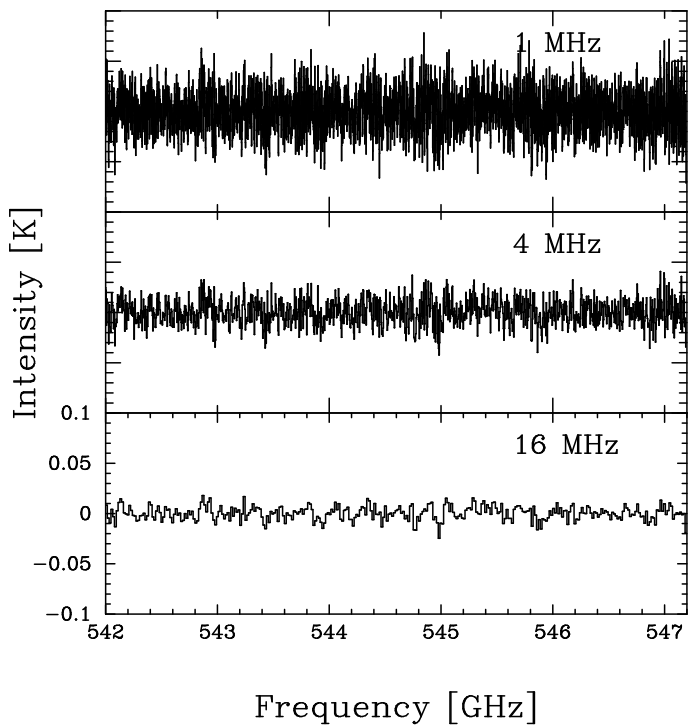}}
 \caption{Examples of the resulting spectra: 2006/07 observations towards
Hot Spot 1 (position A with position B used as reference) is shown with three
different channel  spacings: 1, 4, and 16\,MHz with corresponding 1$\sigma$ noise levels of 20, 13, 7\,mK.}
 \label{figure: AB AOS 3 resolutions}
\end{figure}

The noise is rather stable over the  covered band of all surveys, with a few exceptions where we had loss
of observation time. 
We therefore  measure the  1$\sigma$ noise level across the total  band covered,  towards each position.
We have re-binned the AOS data to the  AC channel spacing of 1\,MHz, and then re-binned both AOS and AC to  channel spacings of 4 and 16\,MHz, 
with an example shown in 
Fig.\,\ref{figure: AB AOS 3 resolutions}.
This rebinning should lower the noise by a factor of 2 and 4,
respectively, if the noise is Gaussian
distributed. However, our measurements  do suffer from baselines at a low level which
prevents the noise to follow Eq.~(\ref{radiometer formula}) exactly.
All 1$\sigma$ results, except the deep searches, are found in Tables~\ref{2004 table noise} and \ref{upper limits with theta}, which also include the peak intensity of the noise given in terms of the \mbox{measured $\sigma$}.

 Figures~\ref{Fig: AB week 1-5 vs. channel} and \ref{Fig: AB week 6-10 vs. channel} (on-line material) show the original AOS spectra used to
produce Fig.\,\ref{figure: AB AOS 3 resolutions}. The intensity is plotted vs. channel numbers and shown for all ten 
frequency settings. The intensity scale in all figures are expressed in terms of observed antenna temperature $T_\mathrm{A}*$.

\emph{Note, that  we do not know if 
the signals we seek are absorption or emission lines,  or if they are located in the signal or  the reference position. }

\subsection{Comments on  results from the 2004 observations}

The observations from summer 2004 resulted in a wide frequency band, but with a rather high noise level. 
In Table~\ref{2004 table noise} we see that 
a typical 1$\sigma$ is about 38\,(85), 23\,(45), and 15\,(26)\,mK for the AOS(AC) observations with 1, 4 and 16\,MHz channel spacing, respectively.

 \begin{table} [!ht]
\caption{Upper limits from observations during 2004 using  1, 4 and 16\,MHz channel spacings. The limits are for the data observed with the AOS(AC) and for frequencies between 547\,--\,563 \mbox{(563\,--\,578)\,GHz}. 
}
\label{2004 table noise} 
\centering
\begin{tabular} {ccc}
\hline
\hline      \noalign{\smallskip}
  \multicolumn{3}{c}{1\,MHz channel spacing}\\\noalign{\smallskip}
Position& 1$\sigma$ & Peak Intensity     \\
& [mK]& [$\sigma$]\\
\noalign{\smallskip}
\hline     \noalign{\smallskip}
HotSpot 1&  37\,(88)    & 3.8\,(4.0) \\
HotSpot 2&  40\,(83)&3.6\,(3.9) \\
     \noalign{\smallskip} \hline   \hline    \noalign{\smallskip} 
  \multicolumn{3}{c}{4\,MHz channel spacing}\\ \noalign{\smallskip}
\hline     \noalign{\smallskip}
HotSpot 1&  22\,(46)  & 3.1\,(3.3) \\
HotSpot 2& 24\,(44)& 3.2\,(3.4) \\
     \noalign{\smallskip} \hline   \hline    \noalign{\smallskip} 
  \multicolumn{3}{c}{16\,MHz channel spacing}\\ \noalign{\smallskip}
\hline     \noalign{\smallskip}
HotSpot 1&  14\,(27) & 2.9\,(3.1)  \\
HotSpot 2& 13\,(25) & 3.0\,(3.2) \\

\hline
\end{tabular}
\end{table}


 \subsection{Comments on results from the 2006/07 observations}

Due to the new scheme
the analysis of these observations is somewhat different from that of the 2004 
observations.  
As seen from Fig.\,\ref{obs strategy 2006-2007}, each pair of observations is sensitive to different angular scales of the perturbations. As an example, if we assume that the signal is in position A and use B as reference, this combination is sensitive to a  perturbation size between \mbox{2\,--\,5\arcmin}. If the primordial cloud is larger than 5\arcmin~the signal will be present in both A and B and will disappear in the resulting AB average. We are, however, also sensitive to larger sizes of the perturbations in A since we have observations towards position D with A as a reference. If we switch and use A as signal and D as reference, we will have an AD average which is sensitive to scales between 2\,--\,30\arcmin. 
Therefore, we can average both AB and AD together to further lower the noise with a sensitivity of scales  2\,--\,5\arcmin.
For larger scales we can only use AD.
A summary of all average combinations   for each position with corresponding angular scale sensitivity is given in  Table~\ref{table +comb. 2nd spectral survey}.

\begin{table} [\!ht]
\caption{Observational strategy 2006/07: different combinations of
 signal and reference positions are sensitive to different angular
sizes of the primordial perturbations.  }
\label{table +comb. 2nd spectral survey} 
\centering
\begin{tabular} {ccccc}
\hline
\hline      \noalign{\smallskip}
Position	& \multicolumn{4}{c}{Angular sensitivity $\theta$}\\
&2\,--\,5\arcmin& 5\,--\,10\arcmin& 10\,--\,15\arcmin	&	15\,--\,30\arcmin   \\
\noalign{\smallskip}
\hline     \noalign{\smallskip}
A& AB+AD & AD&AD&AD\\
B& BC+BA &BC&BC&--\\
C& CD+CB &CD+CB&CB&--\\
D& DA+DC &DA+DC&DA&DA\\
     \noalign{\smallskip} 
\hline
\end{tabular}
\end{table}

This observation strategy is also a way to check if the signal is in the signal or reference position. For example, 
if there is a signal in position A, and the size of the perturbation is  2\,--\,5\arcmin,
the signal-to-noise (S/N) will increase by  $\sqrt{2}$ if we average AB and AD. If the signal is in  B, the signal will be lowered by a factor of two in the AB+AD average.  A  signal in the B position  would also be seen in 
the BC average with  $\sqrt{2}$ increase of the S/N in the BA+BC average.

In Table~\ref{upper limits with theta} we note that the
noise level is considerably
 lower than in the 2004 observations.

\subsection{Comments on results of the deep searches}

The noise level in both deep searches are considerably lower than the 2004 observations and also  lower compared to the surveys in 2006/07. 
The deep search in 2007 resulted in a noise level of 10\,mK with 1\,MHz channel spacing. Rebinning to 4 and 16\,MHz lowers the noise to 6.5, and 3.5\,mK, respectively. The noise level in the
2009 deep search produced an even lower noise level of 6.7\,mK with 1\,MHz channel spacing, and  4.0, and 2.1\,mK when rebinning to
4 and 16\,MHz, respectively.

\begin{figure}[\!ht] 
\centering
\resizebox{\hsize}{!}{\includegraphics{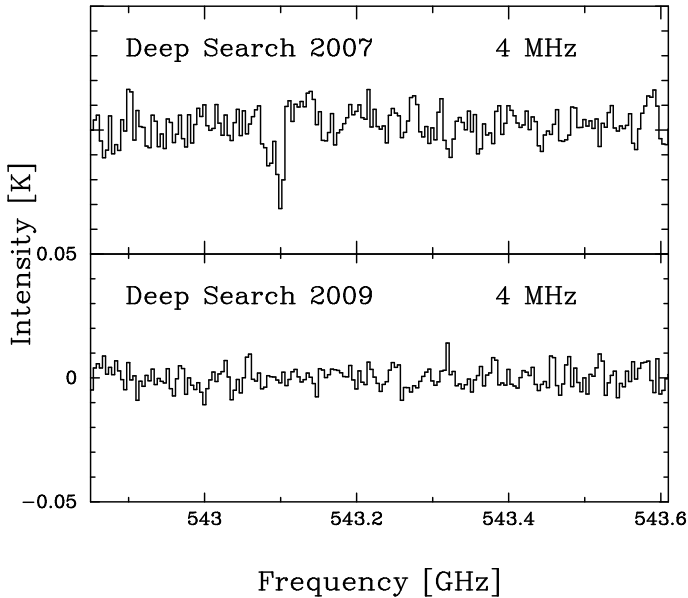}}  
 \caption{Deep searches from 2007 and 2009 observations towards the
Hot Spot 1 (position A with position B used as reference) is shown with  a
channel spacing of 4\,MHz with   1$\sigma$ noise levels of 6.5 and  4.0\,mK, respectively.}
 \label{figure: Line09}
\end{figure}

The possible detection in the 2007 deep search did unfortunately not show up again in the
2009 deep search as seen in Fig.\ref{figure: Line09}. 
The origin of this feature  is difficult to explain but may arise
 from unknown technical issues with the satellite or the receivers.

 \begin{table*} [!ht]
\caption{Upper limits in mK as a function of angular size on the sky from the spectral surveys observations during 2006/07 (deep searches are not included) using  1, 4 and 16\,MHz channel spacings. The limits are for the data taken with the AOS(AC) and for frequency ranges 542.0\,--\,547.5 (486.5\,--\,492)\,GHz.
}
\label{upper limits with theta} 
\centering
\begin{tabular} {ccccccccc}
\hline
\hline      \noalign{\smallskip}
 	&& \multicolumn{6}{c}{1\,MHz channel spacing}\\  \noalign{\smallskip}
Position&2\,--\,5\arcmin& Peak intensity &  5\,--\,10\arcmin& Peak intensity &10\,--\,15\arcmin	&Peak intensity &	15\,--\,30\arcmin &Peak intensity     \\
&1$\sigma$  [mK] &  [$\sigma$] &1$\sigma$ [mK]&   [$\sigma$]&1$\sigma$   [mK]  & [$\sigma$]& 1$\sigma$ [mK]  &  [$\sigma$]\\
\noalign{\smallskip}
\hline     \noalign{\smallskip}
A&16\,(23)&3.0\,(3.3)&  21\,(31) & 3.2\,(3.4)&  21\,(31) & 3.2\,(3.4)&  21\,(31) & 3.2\,(3.4)    \\
B& 17\,(26)&  3.1\,(2.9)&23\,(34)& 3.3\,(3.2)&23\,(34)& 3.3\,(3.2) &--&--\\
C&   19\,(26)&  3.3\,(3.0)& 19\,(26)&  3.3\,(3.0)      &25\,(35)&3.1\,(3.2)&--&--\\
D&15\,(25)&3.1\,(2.9)&  15\,(25) & 3.1\,(2.9) &  20\,(33)  &   3.2\,(3.1)     & 20\,(33)  &   3.2\,(3.1)  \\
     \noalign{\smallskip} \hline   \hline    \noalign{\smallskip}
&&  \multicolumn{6}{c}{4\,MHz channel spacing}\\
\noalign{\smallskip}
\hline     \noalign{\smallskip}
A&10\,(15) &3.0\,(2.8)&  13\,(20) & 3.0\,(2.9)  &13\,(20) & 3.0\,(2.9)      &13\,(20) & 3.0\,(2.9)    \\
B&11\,(17) &2.7\,(3.0)&15\,(23)&2.9\,(3.1)&15\,(23)&2.9\,(3.1)  & --&--\\
C&12\,(16) &3.0\,(2.9)&12\,(16)&3.0\,(2.9)&16\,(22)&2.9\,(3.0) & --&--\\
D&10\,(15) &2.9\,(3.1)&  10\,(15) & 2.9\,(3.1)  &       13\,(20)      &       3.1\,(3.0)       &     13\,(20)    &    3.1\,(3.0)  \\
     \noalign{\smallskip} \hline   \hline    \noalign{\smallskip}
&&  \multicolumn{6}{c}{ 16\,MHz channel spacing}\\
\noalign{\smallskip}
\hline     \noalign{\smallskip}
A&5\,(8)&2.7\,(2.8)& 6\,(11)&3.0\,(2.9)&    6\,(11) &3.0\,(2.9) &6\,(11) &3.0\,(2.9)      \\
B&6\,(9) & 2.8\,(3.0)&   8\,(12)&  2.7\,(3.1) &   8\,(12)&  2.7\,(3.1) &--&--\\
C&6\,(11) & 2.7\,(2.8)& 6\,(11) & 2.7\,(2.8)&      8\,(14)&  3.0\,(2.9) &--&--\\
D&5\,(8)&2.9\,(3.0)& 5\,(8)&2.9\,(3.0)&    7\,(11) &2.8\,(3.1) & 7\,(11) &2.8\,(3.1)    \\

     \noalign{\smallskip} 
\hline
\end{tabular}
\end{table*}

\section{Analysis}

In order to interpret our observations in a cosmological context 
we   begin  
with a summary of the cosmological background and the tools by which we  estimate  the amplitude of absorption   and  resonant lines   in Sect.~\ref{Section: sensitivity analysis}  as well as their line widths. This is 
model dependent 
and we  adopt the currently favoured  hierarchical $\Lambda$CDM   cosmology with a cosmological constant 
in which cold dark matter \mbox{dominates} the evolution of structure.
The latest results from five-years of WMAP data \citep{2009ApJS..180..330K} are used, with a Hubble constant H$_0$\,=\,70\,\kms\,Mpc$^{-1}$,
a dimensionless matter density parameter $\Omega_{\mathrm M}$\,=\,0.274,    a  dark energy density parameter $\Omega_\Lambda$\,=\,0.726, and a baryon density $\Omega_{\mathrm b}$\,=\,0.0456.   This implies a flat geometry where $\Omega_{\mathrm M}$\,+\,$\Omega_\Lambda$\,=\,1.

The proper length $L(z)$ of a primordial perturbation can be estimated by using 
the small-angle limit of the relation between the
angular size $\theta$ that a cloud subtends on the sky and the angular-diameter distance   $D_\mathrm{A}(z)$:
\begin{equation}\label{angular size}
\theta = \frac{L(z)}{D_ \mathrm{A}(z)}\quad[\mathrm{rad}].
\end{equation}
In a flat  Universe,  this distance is described by
\begin{equation}\label{angular diameter distance}
D_\mathrm{A}(z)  = \frac{c}{(1+z\arcmin)H_0} \int_{0}^{z\arcmin} \frac{\mathrm{d} z}{\sqrt{\Omega_{\mathrm M}(1+z)^3+\Omega_\Lambda}}\quad[\mathrm{Mpc}] \ ,
\end{equation}
where $c$ is the speed of light. Assuming a spherical geometry, the  proper length $L(z)$ can be used to calculate the mass  
of a density perturbation $\Delta \rho/\rho$  following the expansion of the Universe
\begin{equation} \label{mass within perturbation}
M = \frac{4\pi}{3}\frac{L(z)^3}{8}\,\Omega_\mathrm{M}\,\rho_\mathrm{cr}(1+z)^3 (1+ \Delta \rho/\rho) \ , 
\end{equation}
where  $\rho_\mathrm{cr}$\,=\,1.88\x10$^{-29}h^2$\,g\,\cmcub~is the critical density of the Universe at present time with $h$\,=\,$H_0$/100,  and  where $\Delta \rho/\rho$ is initially very small but increases with decreasing redshift.
Equations~(\ref{angular size})\,--\,(\ref{mass within perturbation}) then give a relation between angular size and  mass.

The standard structure formation model predicts the formation of gravitationally bound systems from tiny density perturbations via 
gravitational collapse. The hierarchical model predicts that the smaller perturbations formed first
and then merged or accreted gas to form even more massive objects.
The WMAP5 results are consistent with an epoch of reionisation at $z_\mathrm{rei}\!=\!10.9 \pm 1.4$ \citep{2009ApJS..180..330K}. Later 
reionisation epochs at $z\! \sim \!6$, are suggested by other kinds of experiments  \citep[e.g.][]{2006AJ....132..117F}. This implies that the first 
stars must have formed at $z \! \gtrsim \!6$. They are predicted   to form in over-dense dark matter regions of  
\mbox{10$^{5}$\,--\,10$^{6}$\,M$_\odot$}~at  redshifts of about 30\,--\,40
\citep[e.g.][]{2005SSRv..117..445G}. These mini-halos may provide a significant, if not dominant, contribution to reionisation at lower 
redshifts \citep{2008MNRAS.385L..58C}. 
Mass perturbations  $>$10$^{15}$\,M$_\odot$ are  predicted not yet to  have reached their collapse phase. 
 
In the linear regime, the density
contrast  $\Delta\rho/\rho$ is predicted to grow linearly with the scale factor  as
1/(1+z). Sufficiently dense perturbations reach a turn-around point at
which their gravity counterbalances the expansion. Then they enter a
collapse phase during which the density and temperature quickly increase.

\emph{Note that during the linear phase, the only predicted  possible signal   is by the resonant scattering process}, while emission and 
absorption lines will arise in  collapsing perturbations or with high temperature regions for example from the first supernovae and their 
subsequent hot HII regions as background radiation.
The amplitude and line width of all lines depend among many variables on the dynamics of the primordial clouds, which conveniently can be 
divided into the above mentioned phases:
the linear phase, the turn-around phase and the collapse phase.

\subsection{Linear phase:}

In the linear phase, the line width    will depend on the proper length $L(z)$ of the object. This size will occupy a redshift interval $\Delta z$ in the Hubble flow, and, assuming    a spherical   geometry and
that every part of the cloud moves with the same peculiar velocity, we have \citep[e.g.][]{1996ApJ...457....1M, 2008NewA...13...28D}
\begin{equation}\label{fraction of resolution}
\frac{\Delta \nu}{\nu} = \frac{\Delta z}{1+z}  = \frac{H_0\,L(z)}{c} \sqrt{\Omega_{\mathrm M}(1+z)^3+\Omega_\Lambda} \ ,
\end{equation}
where $\Delta \nu$ is the Doppler line width.  The line width is thus dependent on the size of the perturbation.  

A relationship between line width and mass  in the linear phase can now be found using
 Eq.~(\ref{mass within perturbation})
with  $\Delta\rho/\rho \! \ll \!1$ and Eq.~(\ref{fraction of resolution}). 
The line widths increase  with redshift and mass and are very broad,  $\Delta \nu/\nu\!\!\sim$10$^{-1}$\,--\,10$^{-5}$, implying line widths of the order of a few thousand \kms~(Table~\ref{Table: line widths}  on-line material).
 Note, that this is only true assuming that
the density contrast is very small. When the perturbations grow the line widths will start to deviate from Eq.~(\ref{fraction of resolution}) and become increasingly more narrow.
Another addition of uncertainty to the line width is our homogeneity assumption of no substructure within the cloud. 
Within each  mass  at its turn-around, there will be smaller mass perturbations which already have reached their respective turn-around.
The signals from these regions will be very narrow and superposed on the broader lines from the non-collapsed regions. 
As a first order approximation in our analysis, we will therefore use a line width of 500\,\kms~in our predictions of HeH$^+$ resonant lines in the linear phase in Sect.~\ref{Section: sensitivity analysis}.

The intensity of the  resonant  lines is estimated to be
\citep[e.g.][]{1993A&A...269....1D,  1996ApJ...457....1M, 2008NewA...13...28D}
\begin{equation}\label{fractional intensity}
\frac{\Delta I}{I_\mathrm{CMB}} =  (1-e^{-\tau})\,(3-\alpha_\nu) \,\frac{\upsilon_\mathrm{p}}{c} \cos \theta \ ,  
\end{equation}
where $\Delta I$ is the observed intensity, $I_\mathrm{CMB}$ the CMB intensity, 
$\alpha_\nu$ is the spectral index, $\upsilon_\mathrm{p}$ is the peculiar velocity of the perturbation,
$\cos \theta$ is the cosine of the angle with respect to the line of sight of the peculiar velocity, and $\tau$ is the opacity of the transition as calculated in the rest frame of the cloud.
The determining factor resulting in a positive or negative sign of $\Delta I/I$
are the
$\cos \theta$ factor.
The only way to produce resonant scattering "emission" lines is therefore to have a source that moves  along the line of sight towards us, while resonant scattering "absorption" lines will appear when the source moves away from us. Note, that the resonant lines do not require temperature differences, but only a moving cloud, CMB photons and the scattering species.

The spectral index of the CMB $\alpha_\nu\! =\!(\nu/I$)($\mathrm{d}I/\rm{d}\nu)$ is derived in \citet{1996ApJ...457....1M}  where $I_\nu\!=\!B_\nu(T_\mathrm{CMB})$  and has the 
general expression 
\begin{equation}\label{alpha}
\alpha_\nu = 3- \frac{h\,\nu/k T} {1-e^{-h\,\nu/k\,T}}\ .
\end{equation}

The peculiar velocity describes  the motion of   primordial perturbations with respect to the Hubble flow. In the $\Lambda$CDM model these perturbations evolve   due to potential gradients   and increase  with time as \citep{2008gafo.book.....L}
\begin{equation} \label{Eq: peculiar velocity}
\upsilon_\mathrm{p} = \frac{\upsilon_\mathrm{p0}}{\sqrt{1+z}}\ .  
\end{equation}
The most commonly used 
peculiar velocity at present time   for the size of a galaxy cluster is $\upsilon_\mathrm{p0}\!\!\sim$\,600\,\kms, derived from the CMB dipole anisotropy,  which together with our spatially flat $\Lambda$CDM model, is considered to reproduce the characteristics of the large-scale matter distribution at low redshifts  \citep[e.g.][]{1998ApJ...499...20J}. Smaller mass perturbations most likely move with  a factor of 2\,--\,5 higher 
peculiar velocity \citep{2008NewA...13...28D}.
The present ratio of the peculiar velocity to the speed of light   is $\sim$2\x10$^{-3}$ and decreases with increasing redshift. This is a very limiting factor in the search for resonant lines.

The scattering efficiency of resonant lines  is very frequency dependent 
and depends mainly on two parameters:
the number density of the species and the spontaneous transition probability described by  the Einstein coefficient $A_{ul}$, where the 
subscript $ul$ refers to the upper and lower levels of the scattering species. Those species which have a high product of these quantities   are of 
special interest.

The  density of the species  depends on redshift and can    be described 
in terms of the   total density of hydrogen and the fractional abundance of  species $x$ as  $X_x$\,=\,$n_x/n_\mathrm{H}$.
For simplicity, as a first approximation we assume that  the  density is high enough to allow
the
number density of the species   in an excited state  to be  described by the Boltzmann distribution 
\begin{equation} \label{species density} 
n_{x,\mathrm{u}}(z) =     \frac{g_\mathrm{u}}{Q(T)}e^{-E_\mathrm{u}/\mathrm{k}T_\mathrm{ex}}  \,n_ x \ ,
\end{equation}
where $g_\mathrm{u}$ and  $E_\mathrm{u}$  are the statistical weight and energy of the upper state, respectively,  $Q(T)$ is the partition function,
$T_\mathrm{ex}$ is the excitation temperature,   $n_x \!\!=\!\!\,n_ \mathrm{H,0}\, X_{x,0}\, (1+z)^3$ with
$X_{x,0}$ as the abundance extrapolated at the present time   and 
the present density of hydrogen atoms  $n_\mathrm{H,0}=\rho_\mathrm{crit} \Omega_\mathrm{B}/m_\mathrm{p}\!\approx$ 2\x10$^{-7}$\,\cmcub. 
We thus assume that the molecular abundances have reached their asymptotic limit at the redshifts of interest $\lesssim$100\,--\,200 on the large scales
encompassed by the Odin beam. The possibility of a detection prior to a redshift of a maximum abundance is very limited due to the very low abundances and weak signals at even higher redshifts.  
On scales smaller than the Odin beam the abundances will certainly vary in regions with higher than average density or by the effects
of the first stars. 
In reality, the population distribution  also will depend on the molecular formation process,  and the radiation field with possible  population inversion (maser effects).

The optical depth at the centre of the line can be calculated, assuming LTE  and  a Gaussian line profile, using
\begin{equation} \label{tau}
\tau_{\mathrm{max}} =\int_{l_1}^{l_2} \sqrt{\frac{\ln \,2}{16\, \pi^3} }\frac{c^3\,n_x}{ \nu_{{\mathrm{u}l}}^3\,\Delta \upsilon} \,   \,   \frac{A_{{\mathrm{u}l}}\,g_\mathrm{u}}{Q(T)}\,e^{-E_{\mathrm{u}}/kT_\mathrm{ex}} (e^{\,h\nu_{{\mathrm{u}l}}/kT_\mathrm{ex}}-1)\,\mathrm{d}l,
\end{equation}
where we have taken the   stimulated emission into account, and $\nu_{{\mathrm{u}l}}$ is the frequency of the transition \citep[cf.][]{2007A&A...476..807PaperII}.
As customary we have converted the line width in frequency $\Delta \nu$ to a Doppler velocity width $\Delta \upsilon$.
The integration   is performed over the path length of the cloud, $L(z)$\,=\,$l_2-l_1$. Note, that the path length determines the line width in the strictly linear phase (cf. Table~\ref{Table: line widths},  on-line material).

\subsection{Turn-around phase:}

During this phase the tendency of a perturbation to collapse under its
gravity just balances its tendency to expand with the
rest of the Universe.
The perturbation will appear non-moving  and all species from every part of the perturbation will emit from the same redshift.
This   will produce the strongest and the most narrow \emph{resonant} lines with a line width  determined by the thermal broadening \citep{1996ApJ...457....1M}
\begin{equation}\label{thermal broadening}
\frac{\Delta \nu}{\nu} = \frac{2}{c} \sqrt{\frac{2\, \ln2 \,k\,T_\mathrm{K}}{m}} = 7.16\times10^{-7} \sqrt{\frac{T_\mathrm{K}}{A}}\ ,
\end{equation}
where $m$ is the atomic mass and $A$ is the mass in atomic mass units of the species.
At low temperatures this implies a line width of a few \kms~at high redshifts for HeH$^+$  (cf. Table~\ref{Table: line widths},  on-line material).

In addition to this, 
turbulence will also contribute  with an unknown, and perhaps dominating, amplitude. A large region will also have a number of clumps which  have  narrow lines at slightly different velocities.  The total line will be a superposition of  these  lines which will broaden the line. 
Finally, the cloud also consists of a number of smaller clouds that already have reached their turn-around which also may act as broadening agents. As a first order approximation in our analysis, we will therefore use a line width of 30\,\kms~in our predictions of HeH$^+$  resonant lines in the turn-around phase in Sect.~\ref{Section: sensitivity analysis}.

The integrated intensity of the resonant lines has not changed compared to the linear phase, and thus the amplitude, described by Eq.~(\ref{fractional intensity}), will increase proportionally to the decrease
in line width, which could be up to  three orders of magnitude.
 
 As pointed out by \citet{1997A&A...324...27D}, a
luminescence effect  may also produce lines
when  high energy photons are scattered and decay  into several lower energy photons.
The  amplitude of lines produced by luminescence  is given by Eq.~(\ref{fractional intensity}) times an additional 
 gain coefficient factor $K$. 
For simple elastic scattering $K$\,=\,1 and for luminescence $K\!\lesssim\!1000$.
\citet{1995A&A...296..301D} investigated luminescence produced by excited rotational-vibrational (ro-vib) transitions of H$_2$D$^+$ which decay  to the lowest vibrational state   emitting via several rotational transitions.
This effect is, however, limited to  high temperature or high density regions.

\begin{figure}[\!ht] 
\centering
\includegraphics[scale=0.58]{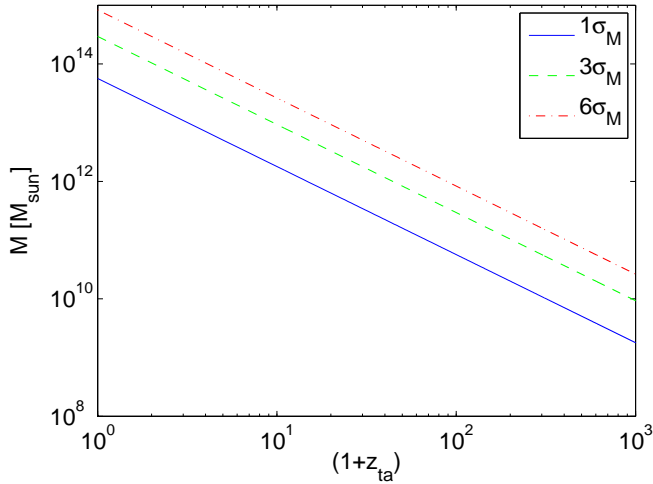}     
\caption{Turn-around redshift for  one, three and six $\sigma_{M}$ mass over-dense primordial perturbations as determined by Eq.~(\ref{turn-around z}). Note, that this describes the total mass, and the
baryonic to dark matter content is 1/6. Star and galaxy formation starts at later times since  the baryonic matter does not collapse immediately at the turn-around redshift. 
}
 \label{Fig: turn-around z vs. mass}
\end{figure}

The turn-around  and the beginning of the collapse phase are  identified as the most favourable for observations of resonant lines  
\citep{1996ApJ...457....1M}.  It is therefore important to try to estimate at what redshift this will take place (full derivation in Appendix~\ref{appendix: derivation TA(z)}, on-line material).
We assume that the power spectrum of the dark matter density fluctuations is of Harrison-Zeldovich type, which means 
\begin{equation} 
\sigma_M = \left \langle \,  \left (\frac{\Delta M}{M} \right)^2 \, \right \rangle^{1/2} \! \! \! \! \sim   M^{\,-2/3}\ ,
\end{equation}
and that it can be normalised by  the observed fluctuations 
\mbox{$\sigma_{M_S}\!=\!10^{-4}$} of the mass $M_S\!\!\!=\!\!\!3.72\times10^{15}$\,M$_\odot$     \citep[cf. Eq. (15.13)][]{2008gafo.book.....L}      
within the  sound horizon at the last scattering surface  where \mbox{$z_\mathrm{\,LSS}\!=\!1\,090$} giving
\begin{equation}
\sigma_{M}=10^{-4}\left(\frac{M_S}{M} \right)^{2/3}\ .
\end{equation}
The turn-around redshift $z_\mathrm{TA}$ for a mass $M$ which has  1\,$\sigma_M$ over density, i.e.
 $\Delta M/M=\sigma_{M}$, can then be estimated according to
\begin{equation}\label{turn-around z}
1+z_{TA} =   1.35\,(1+z_{LSS})\,\left ( \frac{10^9\,M_\odot}{M}  \right)^{2/3}\ .
\end{equation}
Figure~\ref{Fig: turn-around z vs. mass}   shows the turn-around redshift $z$ as a function of mass for one, three, and six $\sigma_M$ mass 
perturbations.
Note, that the Harrison-Zeldovich power spectrum has the power index $n$\,=\,1, and the latest WMAP results  \citep{2009ApJS..180..330K} 
indicate that
\mbox{$n$\,=\,0.960\raisebox{.4ex}{$^{+0.014}_{-0.013}$}}. 
\begin{figure}[\!ht] 
\centering
\includegraphics[scale=0.58]{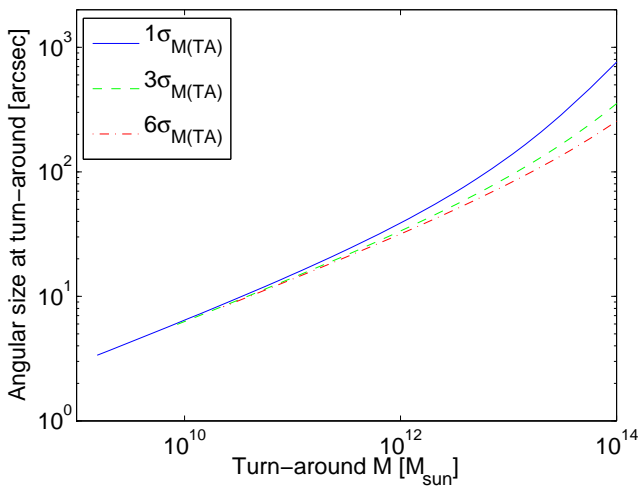}     
\caption{The angular size vs. total mass at $z_\mathrm{TA}$, assuming one, three and six $\sigma_{M}$
over densities.
Note, that this describes the total mass.
}
 \label{Fig: turn-around z vs. angular size}
\end{figure}
In our calculations  we have taken  
the total mass into account, but   the
baryonic to dark matter mass ratio is only 1/6. The star forming baryonic matter is predicted to subsequently fall into the dark matter potential wells  caused by  gravitational collapse. Note, that the collapse does not immediately follow when the object arrives at the turn-around point since the infalling mass has difficulties to loose its energy \citep[e.g.][]{2008ApJ...680L..25D}.
 
To be able to quantify the signals from the proto-objects to observables
we need to relate the turn-around mass to a linear size corresponding to an angular size of our beam.
At the turn-around  $(1+\Delta \rho/\rho)\!\!=\!\!(3\pi/4)^2\!\!\sim\!\!5.55$ and this should be inserted in Eq.~(\ref{mass within perturbation}) to find a mass-linear size relationship at the turn-around. We then use Eq.~(\ref{angular size})\,--\,(\ref{mass within perturbation}) and (\ref{turn-around z})  to find the relation between angular size  vs.  mass perturbations at  turn-around redshift,   
shown in Fig.\,\ref{Fig: turn-around z vs. angular size}.

\subsection{Non-linear collapse phase:}

When (1+ $\Delta \rho/\rho)\!\ga\! (3\pi/4)^2$ the collapse will begin. The amplitude and line width now depend on the collapse rate in addition to the peculiar velocity. Depending on the ratio  of the two velocities resonant lines can appear in absorption, emission or have a double peak  \citep[a summary is found in][]{1996ApJ...457....1M}.  
During the initial stages  the line widths are expected to increase compared to the narrow width during the turn-around. This is also the only 
evolutionary stage
 where the peculiar velocity is no longer required to produce resonant lines.

The chemical abundances could also  be substantially modified during the collapse phase of a primordial cloud when
the increasing temperature and density   induce new chemistry. 
In such regions, molecules will both be destroyed and produced \citep{1996A&A...305..371P}. 
The first Pop III objects will quickly create a  complex, multi-phase interstellar medium with a large range of
densities and temperatures;  up to 6 orders of magnitude at a given radius \citep{2008ApJ...685...40W}.

The very energetic radiation from the first collapsed objects can for instance
dissociate   molecular hydrogen \citep[e.g.][]{1997ApJ...476..458H, 2000ApJ...534...11H}. 
However, if the amount of  H$_2$ produced in the cooling gas behind shock waves from the first  supernova explosions exceeds the destroyed primordial H$_2$ inside the photodissociation regions, the first objects would have a net positive feedback on molecular production and galaxy formation.
\citet{1998ApJ...499L..17F} concludes that multi-supernova explosions propagating through the interstellar medium can produce regions with a very high H$_2$ abundance. For a wide range of physical parameters in such regions he found an H$_2$ 
fraction of  about 6\x10$^{-3}$.
In addition, extremely energetic pair-instability supernovae (140\,--\,260\,M$_\odot$)   produced large amounts of dust at very early times \citep[e.g.][]{2004MNRAS.351.1379S} thereby allowing an increased molecular production including H$_2$ and CO.  
All these effects  influence  the amplitudes of the primordial resonant lines and thermal emission and absorption. 
There are, however, large uncertainties in the complex H$_2$ and HD chemistry \citep{2008MNRAS.388.1627G}.

The resonant scattering will become suppressed 
when the density becomes higher than some critical density depending on species, since frequent collisions will cause  thermal radiation to dominate over resonant scattering    \citep{2007NewAR..51..431B}.
The increasing numbers of collisions will drive
the level populations and the excitation towards the lower (or higher) matter temperature, 
and should enhance the 
spectral line absorption (or emission).
 This effect
will be greatest in molecules like H$_2$ and HD, which have zero or small
dipole moments and thus weak transitions.

  
 %
 
\begin{figure*}[\!ht]
\centering
\subfigure{
\includegraphics[width=.45\textwidth]{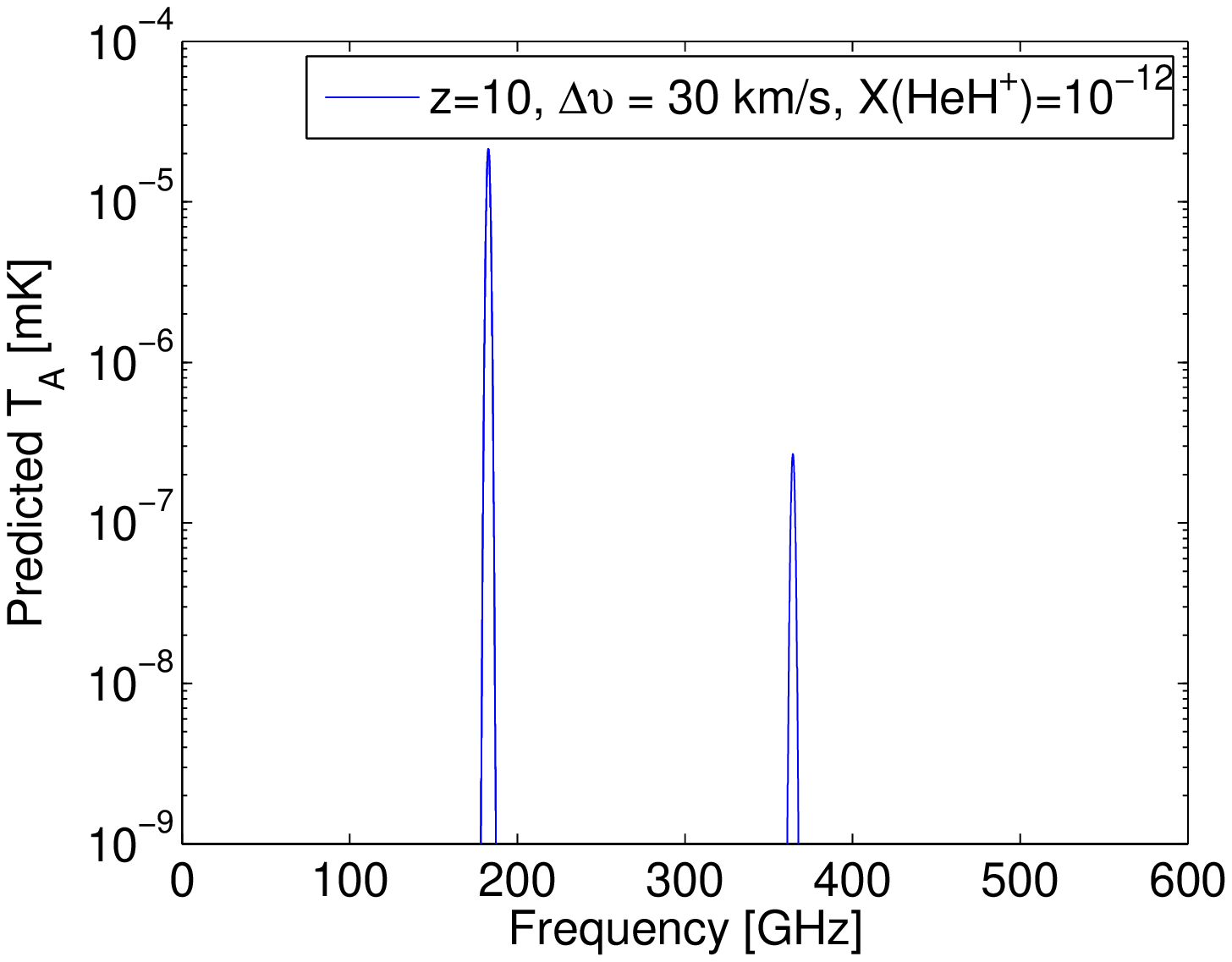}}%
\hspace{.3in}
\subfigure{
\includegraphics[width=.45\textwidth]{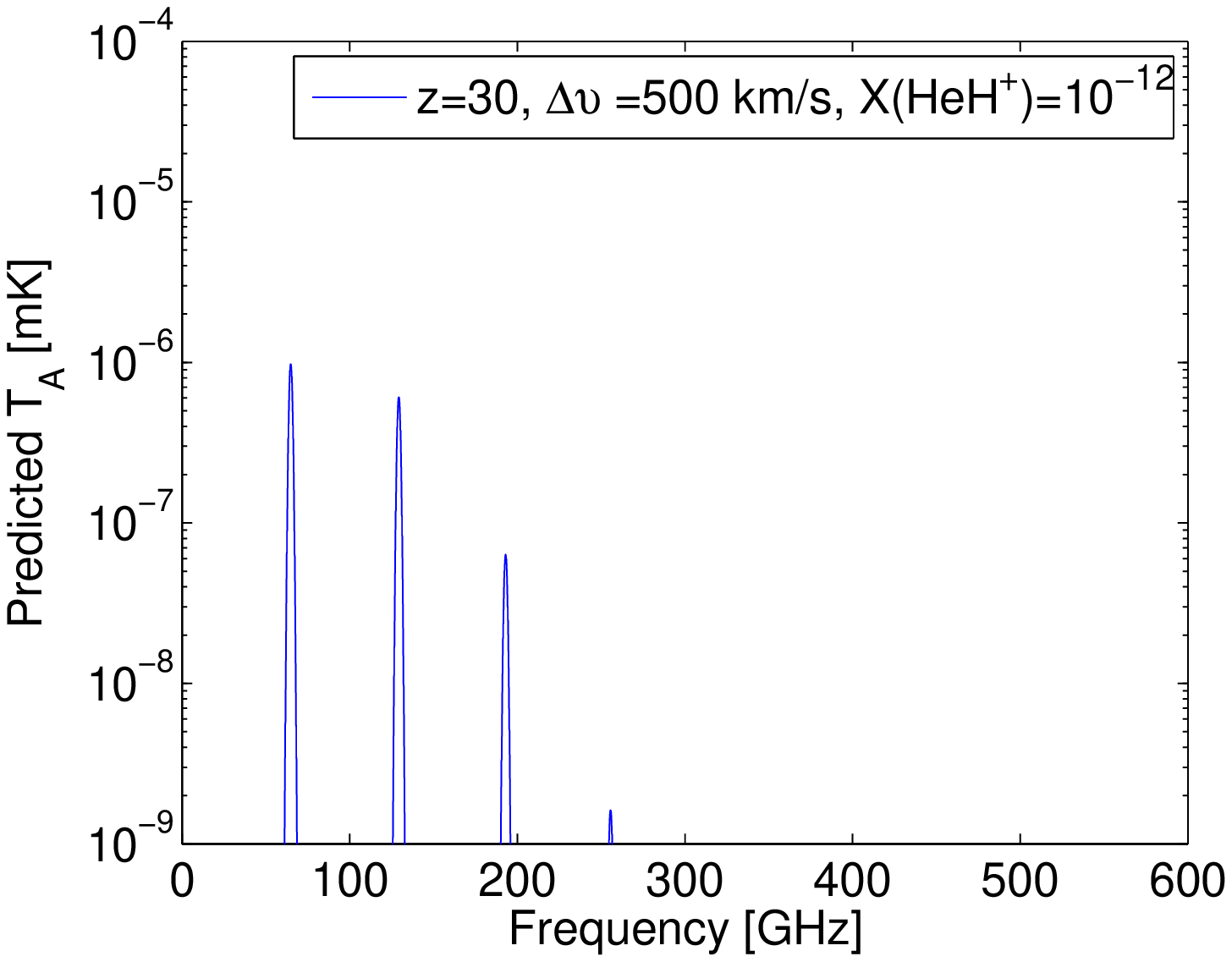}}
\vspace{.3in}
\subfigure{ 
\includegraphics[width=.45\textwidth]{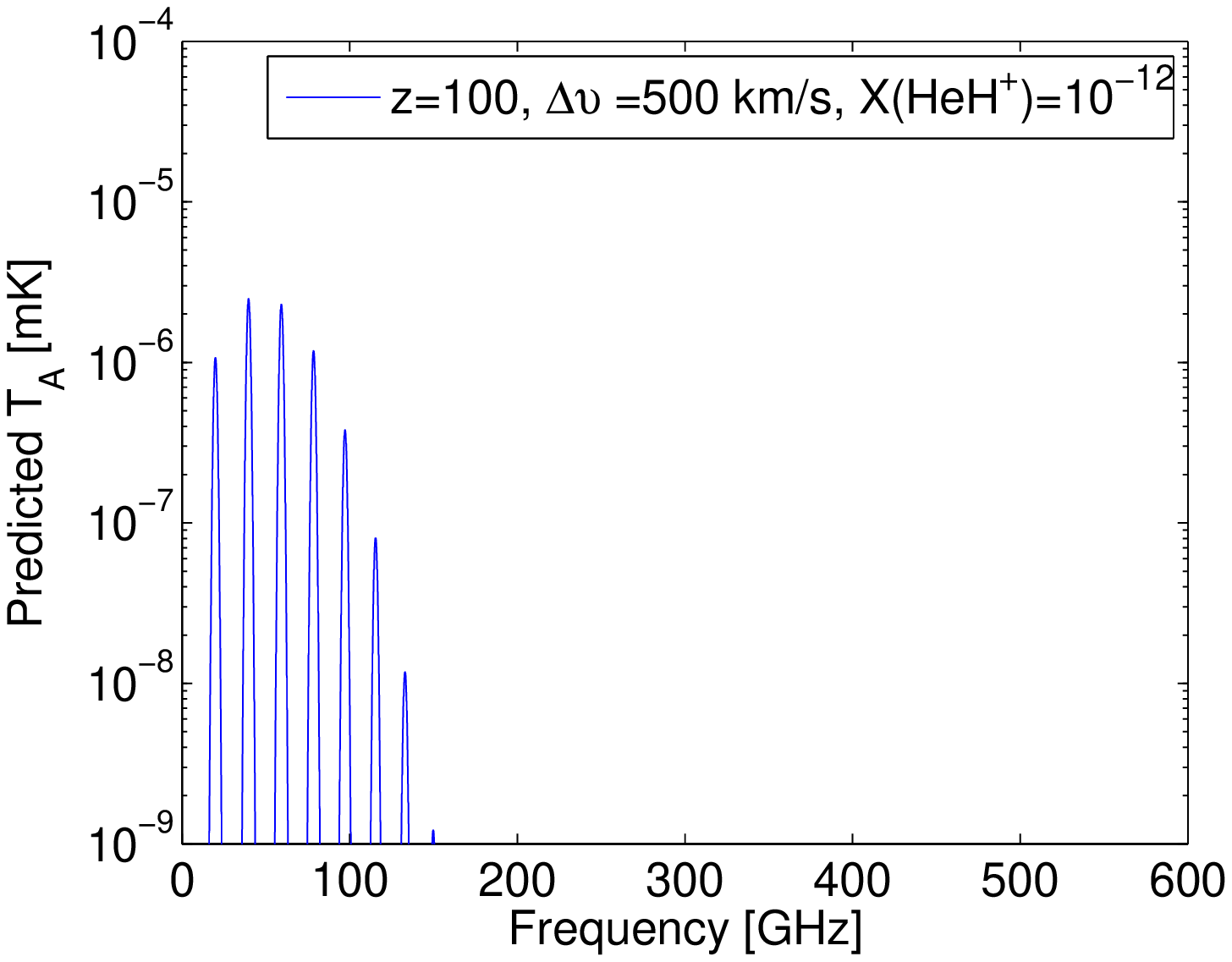}}
\hspace{.3in}
\subfigure{
\includegraphics[width=.45\textwidth]{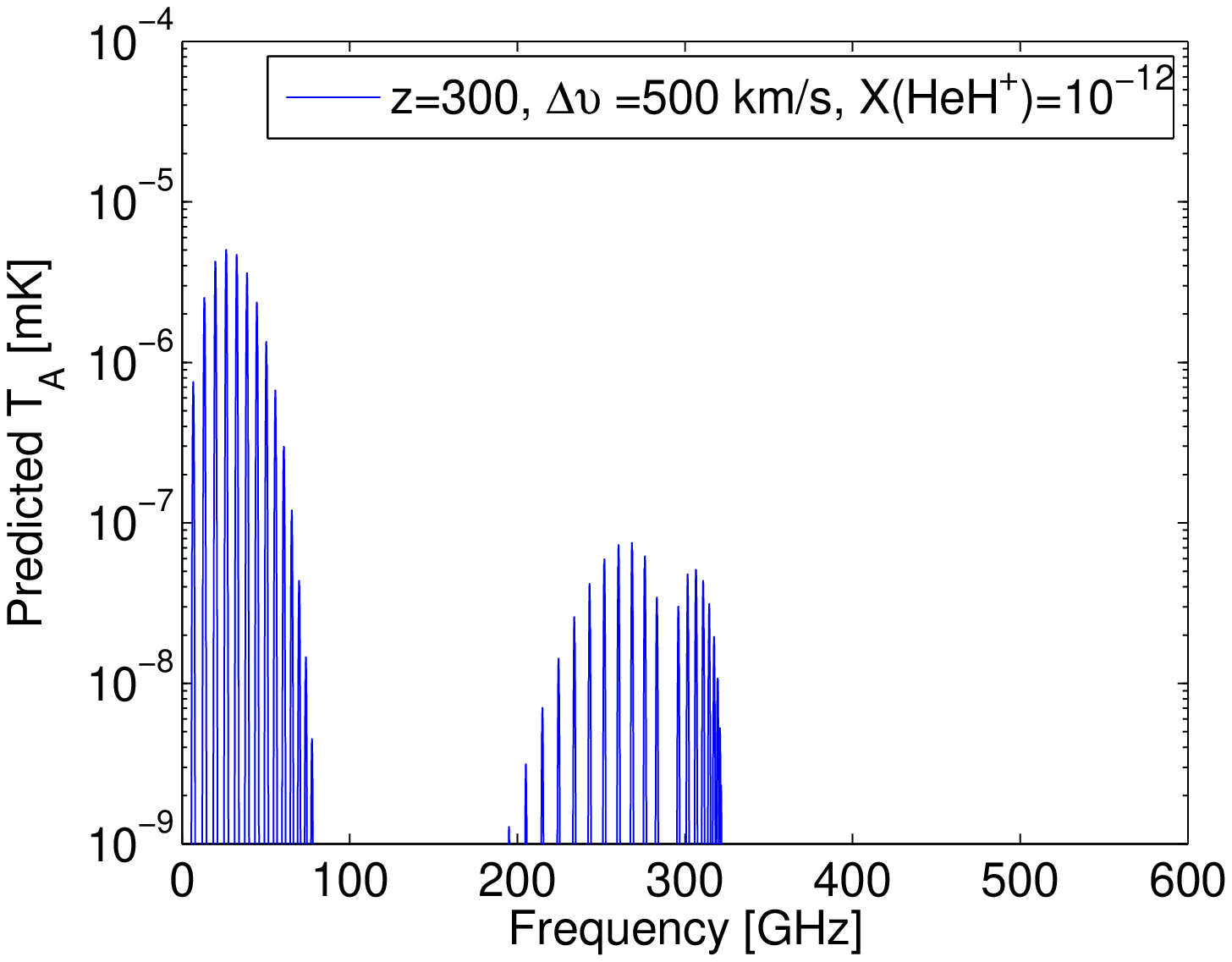}}
\caption{These plots are an example with the main goal to give an idea of the \emph{pattern} of the primordial spectral lines: predicted antenna temperature $\Delta T^*_A$ of resonant    HeH$^+$  optically  thin lines within the $2\farcm1$ Odin beam  at $z$\,=\,10, 30, 100, and 300. Turn-around phase is assumed for 
$z$\,=\,10~and for higher $z$ we assume linear phase with line widths as indicated in the legends.  We  also assume a  low density $n_\mathrm{H}$\,=\,2\x10$^{-7}$\,(1+$z$)$^3$ and 
$X$(HeH$^+$)\,=\,10$^{-12}$. \emph{Note, that any increase of the density, abundance, or decrease of line width,   directly increases the amplitude of the signals with an equal magnitude. If thermal  absorption processes are present, they may be several orders of magnitudes stronger than the resonant lines}. The amplitudes of the ro-vibrational lines are lower than 10$^{-8}$\,mK until $z\!\sim\!300$, when they show up around  200 to 300\,GHz. They are, however, orders of magnitudes weaker than the rotational lines.}
\label{fig: HeH+ lines}
\end{figure*}

\section{Sensitivity analysis}\label{Section: sensitivity analysis}

Our observations have given us upper limits in terms of noise levels, which we now want to analyse 
using Eq.~(\ref{fractional intensity}). This equation is, however,
expressed in intensity while  sub-mm and radio antennas are usually calibrated according to the
brightness temperature in the Rayleigh-Jeans limit in terms of  \citep{2004tra..book.....R}
\begin{equation}\label{T_b}
T_\mathrm{\mathrm{b}} = \frac{c^2}{2\,k\nu^2}\,I_\nu =  \frac{\lambda^2}{2\,k}\,I_\nu \ .
\end{equation}

Thus, the solution of the radiative transport equation  with a constant source function, in our \emph{switched} observations between signal and reference positions, becomes  \citep[cf.][]{2009A&A...494..637P}
\begin{equation}\label{Delta solution}
\Delta T^*_\mathrm{A}   =  T_\mathrm{b}  \, \eta_{\mathrm{mb}}\, \eta_{\mathrm{bf}}= [J(T_\mathrm{ex}) -  J(T_\mathrm{bg})] \, (1-e^{-\tau})  \, \eta_{\mathrm{mb}}\, \eta_{\mathrm{bf}}\ ,
\end{equation} 
where $\Delta T^*_\mathrm{A}$ is the \emph{observed} antenna temperature, $\eta_\mathrm{mb}$ is the main beam efficiency,   $\eta_\mathrm{bf}$ describes the beam filling (see   Eq.~(\ref{Eq: beam-filling}) below), $T_\mathrm{ex}$ and  $T_\mathrm{bg}$ are the excitation and background temperatures, and the 
radiation temperature, $J(T$) is   \citep{2004tra..book.....R}
\begin{equation}\label{radiation temperature}
J(T)  = \frac{h \nu}{k} \frac{1}{e^{\,h\nu/kT}-1} \sim T\ .
\end{equation}
Note that the approximation is only valid   if h$\nu$ $\ll$ k$T$ 
which is often used at frequencies $\lesssim$100\,GHz and  temperatures much higher than the CMB. 
Figure~\ref{Fig: Jx(3-alpha) vs freq} (on-line material) shows  $J(T_\mathrm{CMB})$ 
as a function of frequency with $T$\,=\,2.725\,K.
Around 550\,GHz, the approximation is no longer valid due to the low (CMB) temperature and high frequencies which lowers $J(T_\mathrm{CMB}$) to $\sim$10$^{-3}$.    Accordingly, we use $J(T_\mathrm{ex}$) and not $T_\mathrm{ex}$ in our calculations.
The  5$\sigma$ noise level  from Table~\ref{2004 table noise} and \ref{upper limits with theta} is used as  $\Delta   T^*_\mathrm{A}$ analysing the observations.

The beam filling $\eta_\mathrm{bf}$ can be estimated by \citep[cf.][]{2009A&A...494..637P}
\begin{equation} \label{Eq: beam-filling}
\eta_\mathrm{bf} = \frac{\theta_\mathrm{s}^2}{\theta_\mathrm{s}^2 + \theta_\mathrm{beam}^2}\ , 
\end{equation}
where $\theta_\mathrm{s}$ is the FWHM angular size of the circular source, and $\theta_\mathrm{beam}$ is the FWHM beam size.
If the source is much smaller than the beam we will severely suffer from beam dilution by this factor.
We  therefore assume that the primordial perturbation fills the large Odin beam ($\eta_\mathrm{bf}$\,=\,1)    thereby 
determining the minimum size of a perturbation to which we are sensitive  (Figs.~\ref{Fig: turn-around z vs. mass} and \ref{Fig: turn-around z vs. angular size}). In Table~\ref{Odin step scale} (on-line material) the physical sizes corresponding to the Odin beam as well as the 
step sizes in the 2006/07 observations are given for a number of redshifts. 
According to Figs.~\ref{Fig: turn-around z vs. mass} and \ref{Fig: turn-around z vs. angular size}, 
the Odin beam size is most sensitive to  about 6$\times$10$^{12}$\,--\,10$^{13}$\,M$_\odot$~perturbations with turn-around redshifts of about 3\,--\,10 (1\,--\,3\,$\sigma$).

\subsection{Resonant lines} \label{Subsection: resonant lines sensitivity}

Equation~(\ref{fractional intensity}) gives the intensity of the \emph{resonant} lines. To express this in 
Rayleigh-Jeans (RJ) brightness temperature we use  Eq.~(\ref{Delta solution}) and obtain
\begin{equation}\label{observed Ta}
T_\mathrm{b}  = J(T_\mathrm{CMB})\,  (1-e^{-\tau})\, (3-\alpha_\nu) \, \frac{\upsilon_\mathrm{p}}{c}\cos \theta \ .  
\end{equation}
Note, that  the \emph{observed} antenna temperature is lower by a factor $\eta_\mathrm{bf}\,\eta_\mathrm{mb}$ (Eq.~(\ref{Delta solution})).

To estimate the maximum RJ brightness temperature possible we assume that the lines  in Eq.~(\ref{observed Ta}) are optically \emph{thick} 
($\tau\! >\!\!1$) since this is a limit after which 
the   line will no longer increase its amplitude.
The results are given in
Table.~\ref{Table: max signal}  
with an accuracy 
depending on the uncertainty of the assumed peculiar velocities.
Obviously, even taking this uncertainty into account,  the resonant lines would be extremely difficult to
detect at   frequencies around 500\,GHz and require noise levels orders of magnitudes lower than our observations.  The Odin
main beam temperature   $T_\mathrm{mb}$-scale is related to the flux density by
$F_\nu/T_\mathrm{mb}\!=\!2\,600$~Jy/K at 543\,GHz. 
Figure~\ref{Fig: Jx(3-alpha) vs freq} (on-line material)   shows \mbox{$J(T_\mathrm{CMB})\times (3-\alpha)$}, and demonstrates that we loose almost two orders of magnitude in sensitivity around the observed Odin  frequencies compared to frequencies below 200\,GHz.
This is a very limiting factor for detecting the already   weak resonant lines,  especially  for the
rotational-vibrational transitions which   cannot be observed below 200\,--\,400\,GHz if they    form around $z$\,=\,200\,--\,400. The conclusion is therefore that the rotational-vibrational transitions of resonant lines most likely will be too weak for a detection even with future much more sensitive antennas.
\begin{table} [!ht]
\caption{Estimated largest possible  brightness temperature $ T_\mathrm{b}$ for   \emph{optically thick   resonant} lines,  and $\upsilon_p(z)$\,=\,$600\,/\!\sqrt{1+z}$\,\kms. This velocity is typical for clusters while for a smaller mass it could be higher. 
Note, that these values are the same for \emph{all} species and transitions. }
\centering
\begin{tabular} {c	ccc     }
\hline
\hline
 \noalign{\smallskip}
Obs. Freq.   &  \multicolumn{3}{c}{Maximum   $\mid   T_\mathrm{b} \mid $  [mK]}\\ 
  \noalign{\smallskip}
 [GHz] &	$z$\,=\,10 	 	&$z$\,=\,30	&$z$\,=\,100\\
\noalign{\smallskip}
\hline
   \noalign{\smallskip}	
   1	& 1.6	& 	0.98		&	0.54  	\\
50	&1.5	& 	0.92		&	0.51\\
100	&1.3	& 	0.76		&	0.42	\\
200	&0.64	& 	0.38		&	0.21	\\
300	&0.24	& 	0.14		&	0.078	\\
400	&0.071& 	0.042	&	0.024	\\
500	&0.019& 	0.011	&	0.006	\\
600	&0.005& 	0.003	&	0.002	\\
  \noalign{\smallskip} \hline
\end{tabular}
\label{Table: max signal} 
\end{table}

At  frequencies  $\lesssim$100\,--\,200\,GHz the amplitudes of the resonant lines are orders of magnitudes higher than  around 
500\,--\,600\,GHz,  and hence a detection may be possible if the opacity is high at low frequencies. However, 
the resonant lines suffer from their dependence on 
the peculiar velocity  which lowers their amplitudes by approximately three orders of 
magnitude at $z$\,=\,0,  
and even more at higher redshifts
(Eq.~(\ref{Eq: peculiar velocity}) and (\ref{observed Ta})). Since   thermal emission or absorption lines  do not depend on peculiar velocities, if 
existing, they
have the potential to be much stronger than the resonant lines. 

Note, that
these results also need to be corrected for beam filling, if the object is smaller than the beam, and for the 
beam efficiency to obtain the observable 
antenna temperature. For the Odin satellite this only lowers the amplitude by 10\%, but for other antennas the reduction 
may be more than 50\%.

 
We also estimate the amplitude of optically \emph{thin} resonant lines.  As an example, with the main goal to show the pattern of the resonant lines as a function of redshift and to get a first rough estimate of the amplitudes,  we consider the rotational and first branch of rotational-vibrational \emph{resonant} HeH$^+$ lines for four different redshifts, shown in 
Fig.~\ref{fig: HeH+ lines}. 
A single perturbation is assumed,  with a size filling the   $2\farcm1$ Odin beam. We have
assumed a low density of hydrogen equal to the present one: \mbox{$n_ \mathrm{H}\!=\! 2\times10^{-7} (1+z)^3$} at redshifts $\ge$30, 
 and increased this density   5.55 times at $z\!=\!10$ (density at turn-around). We have used an HeH$^+$  abundance relative to hydrogen of 
$X$(HeH$^+$)\,=\,10$^{-12}$ and the background temperature  follows the CMB at all redshifts. 
The adopted lines widths are narrow  with $\Delta \upsilon$\,=\,30\,\kms~at $z$\,=\,10, and broad  with $\Delta \upsilon$\,=\,500\,\kms~from $z\!\ge\!30$  as indicated in the legends.
This choice of densities and line widths  arises from the fact that for  $z\!\ga\!10$,
the large Odin beam corresponds to much larger physical sizes and masses  than the turn-around objects (Fig.~\ref{Fig: turn-around z vs. mass}), and  at higher redshifts  we are therefore most sensitive to objects in their linear phase. 
The peculiar velocity is described by  Eq.\,(\ref{Eq: peculiar velocity}) with $\upsilon_\mathrm{p0}\!\!\sim$\,600\,\kms.

Since the CMB temperature follows 2.725(1+$z$)\,K, this implies that   high redshift sources have an increased  ability  to populate 
higher molecular energy levels  
since the rising temperatures give the photons   more energy to excite the molecules. This is
shown in Fig.~\ref{fig: HeH+ lines}  where  high molecular  transitions   are increasingly populated towards higher redshifts. The rotational-vibrational transitions are not seen at all until $z \! \sim \!300$, and are always much weaker    than the  rotational lines at all redshifts.
 It is also clearly seen that  lines
observed at frequencies around 500\,GHz    always will be weaker than around 100\,GHz;   a 
clear disadvantage for our current observations. 
This could to some extent  be   compensated by the previously mentioned luminescence process   \citep{1997A&A...324...27D} but   has not 
been considered here.

We should here stress that the predictions of Fig.~\ref{fig: HeH+ lines} are based upon rather conservative
assumptions. Any increase of abundance and density, or decrease of line width, would  directly   increase the
amplitudes of the HeH$^+$ signals. Note also that the
abundance is   a function of redshift and that we here simply have assumed that the abundance has reached a maximum value.  
The HeH$^+$ abundance is indeed very uncertain with a predicted abundance
range of 10$^{-8}$\,--\,10$^{-15}$ \citep{2008NewA...13...28D, 2009A&A...503...47V}. 
 

\subsection{Thermal absorption lines} \label{subsection: Absorption lines sensitivity}

The advantages of the Odin satellite are the absence of adverse effects
from the terrestrial atmosphere and the tunable SSB
receivers, allowing a spectral coverage of a broad band. However, one severe
disadvantage for detection of \emph{resonant} lines with Odin is 
the high   frequencies. 
To detect  resonant scattering lines 
the best observing  frequencies are below 100\,--\,200\,GHz considering the rapid decrease
of the 
radiation temperature   as shown in Fig.\,\ref{Fig: Jx(3-alpha) vs freq} (on-line material).

There are, however, other possibilities of detecting signals from the dark ages. 
\emph{Thermal emission} or \emph{absorption} 
depend  on the competition between radiation and collisions as  given by the radiative transfer equation   
 \begin{equation}\label{solution}
T_\mathrm{b}  =\frac{h \nu_\mathrm{obs}}{k}  \left (\frac{1}{e^{\,h\nu_{\mathrm{u}l}/kT_\mathrm{ex}}-1} - \frac{1}{e^{\,h\nu_{\mathrm{u}l}/kT_\mathrm{bg}}-1} \right)
   \, (1-e^{-\tau}) \ , 
\end{equation}
where  $ T_\mathrm{b}$ is the Rayleigh-Jeans brightness temperature,   $T_\mathrm{ex}$ is the  excitation temperature of a transition  and
$T_\mathrm{bg} $ is the   background   radiation. Note that the latter two temperatures, as well as the opacity,  have to be determined at
the high redshift source, and 
that the  frequencies on the right side in Eq.~\ref{solution}  are not the same.
The excitation temperature depends on the radiation field, which is the CMB at high redshifts, and  at lower redshifts it is dominated by the energetic radiation from the first star formation, and on the densities through collisions. 
The frequency in the first factor is the redshifted rest frequency of the transition which   is the same as the  \emph{observed} frequency, while
the other factors  include the \emph{rest} frequency of the transition at the high redshift source.
 
 Equation~(\ref{solution}) states that
a strong signal requires a large temperature difference between the excitation and background temperatures, which can be produced
by for instance collapsing primordial perturbations  or   the more rapid cooling of the gas temperature compared to the CMB at even higher redshifts. If 
$T_\mathrm{ex} \ll T_\mathrm{bg}$  Eq.~(\ref{solution})  simplifies to
 \begin{equation}\label{absorption solution}
T_\mathrm{b}  = - \frac{h \nu_\mathrm{obs}}{k}   \frac{1}{e^{\,h\,\nu_{\mathrm{u}l}/kT_\mathrm{bg}}-1} 
   \, (1-e^{-\tau}) \ . 
\end{equation}
The maximum Rayleigh-Jeans brightness temperature of  an absorption line can be estimated using Eq.~(\ref{absorption solution}) and 
assuming an
optically thick line, which will cause the last factor to approach unity.    The largest possible   $T_\mathrm{b}$  as a function of observation frequency when the background temperature is the CMB at the corresponding redshift are given in Table~\ref{Table: max absorption}. We have also calculated the largest possible intensity when the excitation temperature is only 10\% and 2\% less than the CMB using Eq.\,(\ref{solution}).  Note, that
to calculate the \emph{observed} antenna temperature the brightness temperature needs to be corrected with the (unknown) beam-filling and beam-efficiency using Eq.~(\ref{Delta solution}).

Also note that
if $T_\mathrm{bg}$\,=\,$T_\mathrm{CMB}(z)$,
the argument of the exponential functions in Eq.~(\ref{absorption solution}) 
can be written as
\begin{equation}
\frac{h\,\nu_{\mathrm{u}l}}{kT_\mathrm{CMB}} = \frac{h\,\nu_{\mathrm{u}l}}{kT_0 (1+z)} = \frac{h \nu_\mathrm{obs}}{k T_0} \ ,
\end{equation}
where $T_0$ is the CMB temperature today.
This implies that the maximum strength of an absorption line is redshift independent assuming an optically thick transition. 
In this case, at a chosen observation frequency the amplitude is   the same for all species and transitions
and is \emph{only}  determined by the known curve $J(T_0)$ in Fig.~\ref{Fig: Jx(3-alpha) vs freq} (on-line material). 

\begin{table} [!ht]
\caption{Estimated  largest possible absorption intensity $ T_\mathrm{b}$  at different observing frequencies for \emph{optically thick     absorption  lines}  assuming that the background temperature is the CMB. We have calculated this using  Eq.~(\ref{solution})  and  $ T_\mathrm{ex}$\,=\,0, $ T_\mathrm{ex}$\,=\,$0.90\times T_\mathrm{CMB}$  and $ T_\mathrm{ex}$\,=\,$0.98\times T_\mathrm{CMB}$. Note that these values are \emph{redshift independent}.}
\centering
\begin{tabular} {c	c   c  c }
\hline
\hline
 \noalign{\smallskip}
Obs. Freq.  &  \multicolumn{3}{c}{Max intensity   $\mid T_\mathrm{b}\mid$}  \\
&$ T_\mathrm{ex}$\,=\,0  & $ T_\mathrm{ex}$\,=\,$0.90\times T_\mathrm{CMB}$  & $ T_\mathrm{ex}$\,=\,$0.98\times T_\mathrm{CMB}$  \\ 
  \noalign{\smallskip}
 [GHz] &	  [mK] &	  [mK] &	  [mK]\\
\noalign{\smallskip}
\hline
   \noalign{\smallskip}
   1	& 2\,700	&	270  & 54  \\
50	& 1\,700	 &	250 		&	51 \\
100	& 1\,000	&	210		& 	42\\
200	& 290	&	97		&	21\\
300	& 73	&	33		&	7.5\\
400	& 17	&	9.1		&	2.2\\
500	& 3.6	&	2.2		&	0.6\\
600	& 0.74	&	0.5		&	0.1\\
  \noalign{\smallskip} \hline
\end{tabular}
\label{Table: max absorption} 
\end{table}

If the background temperature is not the CMB, the argument of the exponential function then becomes
\begin{equation}
\frac{h\,\nu_{\mathrm{u}l}}{kT_\mathrm{bg}}   = \frac{h \nu_\mathrm{obs} (1+z)}{k T_\mathrm{bg}} \ .
\end{equation}
As seen, 
the maximum absorption strength in this case depends on both the redshift and the  unknown  background temperature.

\begin{table} [!ht] 
\caption{Estimated required background temperatures to produce absorption lines with 
\mbox{$T_\mathrm{b}=-10$\,mK}  at an observing frequency of  550\,GHz assuming  \mbox{$T_\mathrm{ex} \ll T_\mathrm{bg}$}, and an opacity of one or 10$^{-3}$.
Note, that since the observing frequencies are the same for all transitions the source redshift is different for every transition. As a comparison, the CMB temperatures are listed for each redshift.} 
\centering
\begin{tabular} {lccrr}
\hline
\hline
 \noalign{\smallskip}
Transition & Redshift  & Opacity &  	$T_\mathrm{bg}$  &  $T_\mathrm{CMB}$\\ 
\noalign{\smallskip}
  & $z$   & $\tau$ & [K]& [K]\\   
  \noalign{\smallskip}
\hline
   \noalign{\smallskip}

H$_2$(2\,--\,0)	&	18.3		&	1			&	58	&	53\\  
				&			&	10$^{-3}$	&	260 \\   
H$_2$(3\,--\,1)	& 	31.0		&	1			&	96		& 87\\  
				&			&	10$^{-3}$	&	430 \\ 
H$_2$(4\,--\,2)	& 	43.4		&	1			&	132	&	121	\\ 
				&			&	10$^{-3}$	&	595 \\
H$_2$(5\,--\,3)	& 	 55.4	&	1			&	168		& 154 \\ 
				&			&	10$^{-3}$	&	756 \\			

\noalign{\smallskip}
 HD(4\,--\,3)		&	18.1		&	1			&	57	&	52	\\  
			        &			&	10$^{-3}$	&	257 \\   
  
HD(5\,--\,4)		&	22.7		&	1			&	71		&	65 \\  
			        &			&	10$^{-3}$	&	317 \\    
\noalign{\smallskip}
   
HeH$^+$(4\,--\,3)	&	13.4		&	1			&	 43	&	39	\\  
			        &			&	10$^{-3}$	&	  194&	 	\\    
   
HeH$^+$(5\,--\,4)	&	16.7		&	1			&	 54	&	48\\  
			        &			&	10$^{-3}$	&	240	&  \\

  \noalign{\smallskip} \hline
\end{tabular}
\label{Table: required Tbg for TA 10mK} 
\end{table}

The  noise levels in our observations can be used to set upper limits on possible absorption lines at the observed 
frequencies. The 
background source  cannot, however, be 
the CMB at these frequencies since the maximum possible absorption intensity of such a signal is about 3\,mK 
(Table\,\ref{Table: max absorption}).   
This corresponds to about 1$\sigma$ of our best observations (deep searches), and even less for the spectral line surveys. Thus,  the analysis in order to put upper limits using the Odin 
observations depends on both  the 
unknown background temperature and the redshift, in addition to the opacity and excitation temperature.
We give a very simple example of such an analysis in Table\,\ref{Table: required Tbg for TA 10mK}.  We 
have here estimated the required 
background 
temperature  (other than the CMB) in order to produce an absorption line with $T_\mathrm{b}$\,=\,10\,mK at 
an observing frequency of 550\,GHz for some molecular transitions. We have used two different opacities   
and assumed that  $T_\mathrm{ex} \ll T_\mathrm{bg}$ (Eq.\,(\ref{absorption solution})).  Such an absorption 
line would correspond to about a   5\,$\sigma$ detection for our 2009 deep search using 16\,MHz channel spacing, or a 1\,--\,2\,$\sigma$ detection for our line surveys using  16\,MHz channel spacings (Table~\ref{upper limits with theta}).    We have also listed the CMB temperature at the 
corresponding redshift as a 
comparison.  
A   more 
sophisticated modelling, including excitation temperatures and estimated opacities,  is demanded to produce more realistic results and to investigate how the intensities of different transitions vary with the physical conditions at different epochs.

\section{Summary}\label{section summary}

In order to constrain cosmological models of star  and structure formation as well as the chemical evolution in the early Universe,  
we have performed  spectral line surveys towards several positions without any known sources of emission in a search for primordial spectral lines from the Dark Ages.
The first survey covered a broad band of 31\,GHz between 547\,--\,578\,GHz towards two positions with fixed reference-positions. The second survey covered 11\,GHz in the bands
542.0\,--\,547.5\,GHz and 486.5\,--\,492.0\,GHz towards four positions.
We also performed two  deep searches towards one position with 543.100 and 543.250\,GHz as centre frequencies.
No lines were detected, and thus the results are upper limits in terms of noise level. Typical 1$\sigma$ values  are \mbox{5\,--\,35\,mK} in the 11\,GHz survey,  \mbox{14\,--\,90\,mK} in the 31\,GHz  survey, and 2\,--\,7\,mK in the deep searches over a 1\,GHz band.

The major improvement  made by the Odin observations compared to \citet{1993A&A...269....1D}
is  the  broad bandwidth covered   allowing a wide range of redshifts to be explored  for a number of atomic and molecular species. 
In addition,  in the second survey we have taken into account the unknown sizes of the clouds by
testing an observational strategy 
where we have observed towards four positions in a sequence with different angular distances between the reference
and signal position. 
An important benefit of our observations is that we do no
suffer from spectral line contamination from the terrestrial atmosphere.

At low densities and in matter-radiation equilibrium conditions, the \emph{only}  expected signal
is from resonant line scattering between CMB photons and matter moving with respect to the expansion of the Universe. These lines suffer, however, from the dependence of the low CMB temperature and the peculiar velocities of the moving primordial perturbations which is about 10$^{-3}$ at present and even lower at higher redshifts.
In order to obtain an estimate of the \emph{highest possible intensities} we assume  optically thick resonant lines and no beam-dilution. The intensities will then be at most   of the order of a few mK  at   frequencies below 100\,--\,200\,GHz, and orders of magnitude lower at higher frequencies (5\,$\mu$K at 600\,GHz). Since the lines most likely have $\tau\!\ll\!1$ this will further lower their intensities with orders of magnitudes.

If existing, thermal absorption lines on the other hand have the potential to be 
observable at both high and low frequencies. Such lines do not suffer from the dependence of peculiar velocities of the moving primordial perturbations. The background radiation could also be considerably higher than the CMB, thereby producing stronger lines. The background could for instance be collapsing primordial perturbations, the first stars which are predicted to form at z\,$\sim$\,20\,--\,30, or the remnants of pop~III supernovae. These objects
probably emitted large amounts of energetic radiation which foreground primordial perturbations could absorb.
Also here we find the
\emph{highest possible intensities}  by  assuming optically thick lines  and no beam-dilution. 
Assuming  that the CMB is the background radiation, we find that the largest possible intensity   is 1\,--\,2\,K 
below 100\,GHz, and 
of the order of a few mK around 500\,GHz. If the background radiation is higher than the CMB, the intensities will 
increase. Also in this case the opacity  most likely is $\ll\!\!1$ which will lower the absorption line intensities, in 
addition to beam-dilution.

The strength and line-width of the primordial lines depend on the evolution of the primordial perturbations. In 
the first (linear) phase they are very broad and weak, but become increasingly stronger and more narrow during 
the evolution of the cloud.  The turn-around phase, which produces the strongest and most narrow lines, and the beginning of the collapse have previously 
been identified as  the most favourable evolutionary phases for observations
 \citep{1996ApJ...457....1M}.
We have made a simple estimation of the redshifts and sizes of the  primordial perturbations at their turn-around 
phase, and find 
that the Odin beam size of 2$\farcm$1 corresponds to  a  turn-around  mass of about 
6$\times$10$^{12}$\,M$_\odot$ at a corresponding redshift of about 3 at a 1$\sigma$ level. Smaller perturbations at higher 
redshift will thus suffer from beam-dilution. 
A   beam-size of  40\arcsec, which is approximately the beam-size of Herschel 
Space Observatory\footnote{\url{http://herschel.esac.esa.int/}},  corresponds to a 1$\sigma$ turn-around mass and redshift of about 10$^{12}$\,M$_\odot$ and $z\!
\sim\!10$.

The lowest rotational transitions of H$_2$ will fall in the Odin band around 500\,GHz from z$\sim$20\,--\,30, 
while the lowest transitions of other species such as HD, HD$^+$, and HeH$^+$ will fall in the band around 
100\,GHz from the same epoch. These transitions may be searched for using ground based antennas -- even 
though the foreground and atmospheric radiation will pose a problem.

An important aspect of our work has been to test different
observational strategies to prepare for our forthcoming observations with the much more
sensitive telescope and receivers aboard  the Herschel
Space Observatory  launched on
May 14, 2009.  The much lower noise level and the even broader band coverage with Herschel will increase the 
possibility of a detection.  Other interesting facilities to consider in the searches for primordial molecules are for 
example 
the Atacama
Large Millimeter Array\footnote{\url{http://www.alma.info/}} (ALMA), 
the
Combined Array for Research in Millimeter-Wave 
Astronomy\footnote{\url{http://www.mmarray.org/}}~(CARMA),    the IRAM Plateau de Bure 
Interferometer and the IRAM 30-m telescope \footnote{\url{http://www.iram.fr/}},  and
the Very Large Array\footnote{\url{http://www.vla.nrao.edu/}} (VLA) at 20\,--\,45\,GHz.
Our observing methods and resulting   limits,   
paired with a sensitivity analysis taking into account the evolution of primordial perturbations, should be a
valuable input to the planning of these observations.

Spectral lines from primordial atoms and molecules   may very well be the only way to probe the epoch of the cosmic Dark Ages and its end
when the first stars formed.
The search for    these primordial signals   is a very difficult task indeed, but a detection could be possible with the use of future facilities and would introduce an additional important way to discriminate between models of the early universe as well as star and structure formation.

\begin{acknowledgements}
Research on resonant primordial lines started in our collaboration by 
Francesco Melchiorri nearly twenty years ago. Now that he left us we miss 
his enthusiasm and creativity and we would like to dedicate  this work to 
his memory.
Many thanks also to John~H.~Black and Per Bergman for valuable comments and discussions, and to the whole
Odin team. 
We also thank the anonymous referee whose constructive comments led to an improvement of the paper.
Generous financial support from the Research Councils and Space Agencies in Sweden, Canada, Finland and France is gratefully acknowledged.

\end{acknowledgements}

\bibliographystyle{aa-package/bibtex/aa}
\bibliography{references}

\begin{thebibliography}{95}
\expandafter\ifx\csname natexlab\endcsname\relax\def\natexlab#1{#1}\fi

\bibitem[{{Abel} {et~al.}(2000){Abel}, {Bryan}, \&
  {Norman}}]{2000ApJ...540...39A}
{Abel}, T., {Bryan}, G.~L., \& {Norman}, M.~L. 2000, \apj, 540, 39

\bibitem[{{Applegate} {et~al.}(1987){Applegate}, {Hogan}, \&
  {Scherrer}}]{1987PhRvD..35.1151A}
{Applegate}, J.~H., {Hogan}, C.~J., \& {Scherrer}, R.~J. 1987, \prd, 35, 1151

\bibitem[{{Barkana} \& {Loeb}(2001)}]{2001PhR...349..125B}
{Barkana}, R. \& {Loeb}, A. 2001, \physrep, 349, 125

\bibitem[{{Basu}(2007)}]{2007NewAR..51..431B}
{Basu}, K. 2007, New Astronomy Review, 51, 431

\bibitem[{{Basu} {et~al.}(2004){Basu}, {Hern{\'a}ndez-Monteagudo}, \&
  {Sunyaev}}]{2004A&A...416..447B}
{Basu}, K., {Hern{\'a}ndez-Monteagudo}, C., \& {Sunyaev}, R.~A. 2004, \aap,
  416, 447

\bibitem[{{Baugh}(2006)}]{2006RPPh...69.3101B}
{Baugh}, C.~M. 2006, Reports on Progress in Physics, 69, 3101

\bibitem[{{Becker} {et~al.}(2001){Becker}, {Fan}, {White}, {Strauss},
  {Narayanan}, {Lupton}, {Gunn}, {Annis}, {Bahcall}, {Brinkmann}, {Connolly},
  {Csabai}, {Czarapata}, {Doi}, {Heckman}, {Hennessy}, {Ivezi{\'c}}, {Knapp},
  {Lamb}, {McKay}, {Munn}, {Nash}, {Nichol}, {Pier}, {Richards}, {Schneider},
  {Stoughton}, {Szalay}, {Thakar}, \& {York}}]{2001AJ....122.2850B}
{Becker}, R.~H., {Fan}, X., {White}, R.~L., {et~al.} 2001, \aj, 122, 2850

\bibitem[{{Bertoldi} {et~al.}(2003){Bertoldi}, {Cox}, {Neri}, {Carilli},
  {Walter}, {Omont}, {Beelen}, {Henkel}, {Fan}, {Strauss}, \&
  {Menten}}]{2003A&A...409L..47B}
{Bertoldi}, F., {Cox}, P., {Neri}, R., {et~al.} 2003, \aap, 409, L47

\bibitem[{{Black}(2006)}]{2006RoyalSocietyofChemistryB}
{Black}, J.~H. 2006, in Faraday Discussions of the Royal Society of Chemistry
  (UK), Vol. 133, Chemistry and cosmology, 27--32

\bibitem[{{Bovino} {et~al.}(2009){Bovino}, {Wernli}, \&
  {Gianturco}}]{2009ApJ...699..383B}
{Bovino}, S., {Wernli}, M., \& {Gianturco}, F.~A. 2009, \apj, 699, 383

\bibitem[{{Bowman} {et~al.}(2009){Bowman}, {Morales}, \&
  {Hewitt}}]{2009ApJ...695..183B}
{Bowman}, J.~D., {Morales}, M.~F., \& {Hewitt}, J.~N. 2009, \apj, 695, 183

\bibitem[{{Bromm} \& {Larson}(2004)}]{2004ARA&A..42...79B}
{Bromm}, V. \& {Larson}, R.~B. 2004, \araa, 42, 79

\bibitem[{{Bromm} {et~al.}(2009){Bromm}, {Yoshida}, {Hernquist}, \&
  {McKee}}]{2009Natur.459...49B}
{Bromm}, V., {Yoshida}, N., {Hernquist}, L., \& {McKee}, C.~F. 2009, \nat, 459,
  49

\bibitem[{{Campos} {et~al.}(2007){Campos}, {Saucedo Morales}, {Lipovka}, \&
  {Nunes-Lopez}}]{2007IAUS..235..413C}
{Campos}, J.~C., {Saucedo Morales}, J.~C., {Lipovka}, A.~A., \& {Nunes-Lopez},
  R. 2007, in IAU Symposium, ed. F.~{Combes} \& J.~{Palous}, Vol. 235, 413--413

\bibitem[{{Cherchneff} \& {Dwek}(2009)}]{2009ApJ...703..642C}
{Cherchneff}, I. \& {Dwek}, E. 2009, \apj, 703, 642

\bibitem[{{Cherchneff} \& {Lilly}(2008)}]{2008ApJ...683L.123C}
{Cherchneff}, I. \& {Lilly}, S. 2008, \apjl, 683, L123

\bibitem[{{Chluba} {et~al.}(2007){Chluba}, {Rubi{\~n}o-Mart{\'{\i}}n}, \&
  {Sunyaev}}]{2007MNRAS.374.1310C}
{Chluba}, J., {Rubi{\~n}o-Mart{\'{\i}}n}, J.~A., \& {Sunyaev}, R.~A. 2007,
  \mnras, 374, 1310

\bibitem[{{Chluba} \& {Sunyaev}(2006)}]{2006A&A...458L..29C}
{Chluba}, J. \& {Sunyaev}, R.~A. 2006, \aap, 458, L29

\bibitem[{{Choudhury} {et~al.}(2008){Choudhury}, {Ferrara}, \&
  {Gallerani}}]{2008MNRAS.385L..58C}
{Choudhury}, T.~R., {Ferrara}, A., \& {Gallerani}, S. 2008, \mnras, 385, L58

\bibitem[{{Ciardi} \& {Ferrara}(2005)}]{2005SSRv..116..625C}
{Ciardi}, B. \& {Ferrara}, A. 2005, Space Science Reviews, 116, 625

\bibitem[{{de Bernardis} {et~al.}(1993){de Bernardis}, {Dubrovich}, {Encrenaz},
  {Maoli}, {Masi}, {Mastrantonio}, {Melchiorri}, {Melchiorri}, {Signore}, \&
  {Tanzilli}}]{1993A&A...269....1D}
{de Bernardis}, P., {Dubrovich}, V., {Encrenaz}, P., {et~al.} 1993, \aap, 269,
  1

\bibitem[{{de Bernardis} {et~al.}(1990){de Bernardis}, {Masi}, {Melchiorri}, \&
  {Melchiorri}}]{1990ApJ...357....8D}
{de Bernardis}, P., {Masi}, S., {Melchiorri}, B., \& {Melchiorri}, F. 1990,
  \apj, 357, 8

\bibitem[{{De Lucia} \& {Poggianti}(2008)}]{2008ASPC..399..314D}
{De Lucia}, G. \& {Poggianti}, B.~M. 2008, in Astronomical Society of the
  Pacific Conference Series, ed. {T.~Kodama, T.~Yamada, \& K.~Aoki}, Vol. 399,
  314

\bibitem[{{Diemand} \& {Kuhlen}(2008)}]{2008ApJ...680L..25D}
{Diemand}, J. \& {Kuhlen}, M. 2008, \apjl, 680, L25

\bibitem[{{Dubrovich} {et~al.}(2008){Dubrovich}, {Bajkova}, \&
  {Khaikin}}]{2008NewA...13...28D}
{Dubrovich}, V., {Bajkova}, A., \& {Khaikin}, V.~B. 2008, New Astronomy, 13, 28

\bibitem[{{Dubrovich}(1977)}]{1977SvAL....3..128D}
{Dubrovich}, V.~K. 1977, Soviet Astronomy Letters, 3, 128

\bibitem[{{Dubrovich}(1997)}]{1997A&A...324...27D}
{Dubrovich}, V.~K. 1997, \aap, 324, 27

\bibitem[{{Dubrovich} \& {Lipovka}(1995)}]{1995A&A...296..301D}
{Dubrovich}, V.~K. \& {Lipovka}, A.~A. 1995, \aap, 296, 301

\bibitem[{{Dunkley} {et~al.}(2009){Dunkley}, {Komatsu}, {Nolta}, {Spergel},
  {Larson}, {Hinshaw}, {Page}, {Bennett}, {Gold}, {Jarosik}, {Weiland},
  {Halpern}, {Hill}, {Kogut}, {Limon}, {Meyer}, {Tucker}, {Wollack}, \&
  {Wright}}]{2009ApJS..180..306D}
{Dunkley}, J., {Komatsu}, E., {Nolta}, M.~R., {et~al.} 2009, \apjs, 180, 306

\bibitem[{{Ellis} \& {Silk}(2007)}]{2007arXiv0712.2865E}
{Ellis}, R. \& {Silk}, J. 2007, arXiv: 0712.2865

\bibitem[{{Fan} {et~al.}(2006){Fan}, {Strauss}, {Becker}, {White}, {Gunn},
  {Knapp}, {Richards}, {Schneider}, {Brinkmann}, \&
  {Fukugita}}]{2006AJ....132..117F}
{Fan}, X., {Strauss}, M.~A., {Becker}, R.~H., {et~al.} 2006, \aj, 132, 117

\bibitem[{{Ferrara}(1998)}]{1998ApJ...499L..17F}
{Ferrara}, A. 1998, \apjl, 499, L17

\bibitem[{{Fixsen} {et~al.}(1996){Fixsen}, {Cheng}, {Gales}, {Mather},
  {Shafer}, \& {Wright}}]{1996ApJ...473..576F}
{Fixsen}, D.~J., {Cheng}, E.~S., {Gales}, J.~M., {et~al.} 1996, \apj, 473, 576

\bibitem[{{Frisk} {et~al.}(2003){Frisk}, {Hagstr{\"o}m}, {Ala-Laurinaho},
  {Andersson}, {Berges}, {Chabaud}, {Dahlgren}, {Emrich}, {Flor{\'e}n},
  {Florin}, {Fredrixon}, {Gaier}, {Haas}, {Hirvonen}, {Hjalmarsson},
  {Jakobsson}, {Jukkala}, {Kildal}, {Kollberg}, {Lassing}, {Lecacheux},
  {Lehikoinen}, {Lehto}, {Mallat}, {Marty}, {Michet}, {Narbonne}, {Nexon},
  {Olberg}, {Olofsson}, {Olofsson}, {Orign{\'e}}, {Petersson}, {Piironen},
  {Pons}, {Pouliquen}, {Ristorcelli}, {Rosolen}, {Rouaix}, {R{\"a}is{\"a}nen},
  {Serra}, {Sj{\"o}berg}, {Stenmark}, {Torchinsky}, {Tuovinen}, {Ullberg},
  {Vinterhav}, {Wadefalk}, {Zirath}, {Zimmermann}, \&
  {Zimmermann}}]{2003A&A.402.27.Frisk.etal}
{Frisk}, U., {Hagstr{\"o}m}, M., {Ala-Laurinaho}, J., {et~al.} 2003, \aap, 402,
  L27

\bibitem[{{Frye} {et~al.}(2008){Frye}, {Bowen}, {Hurley}, {Tripp}, {Fan},
  {Holden}, {Guhathakurta}, {Coe}, {Broadhurst}, {Egami}, \&
  {Meylan}}]{2008ApJ...685L...5F}
{Frye}, B.~L., {Bowen}, D.~V., {Hurley}, M., {et~al.} 2008, \apjl, 685, L5

\bibitem[{{Furlanetto} {et~al.}(2009){Furlanetto}, {Lidz}, {Loeb}, {McQuinn},
  {Pritchard}, {Alvarez}, {Backer}, {Bowman}, {Burns}, {Carilli}, {Cen},
  {Cooray}, {Gnedin}, {Greenhill}, {Haiman}, {Hewitt}, {Hirata}, {Lazio},
  {Mesinger}, {Madau}, {Morales}, {Oh}, {Peterson}, {Pihlstr{\"o}m}, {Shapiro},
  {Tegmark}, {Trac}, {Zahn}, \& {Zaldarriaga}}]{2009astro2010S..83F}
{Furlanetto}, S.~R., {Lidz}, A., {Loeb}, A., {et~al.} 2009, Astronomy, 2010, 83

\bibitem[{{Galli} \& {Palla}(1998)}]{1998A&A...335..403G}
{Galli}, D. \& {Palla}, F. 1998, \aap, 335, 403

\bibitem[{{Galli} \& {Palla}(2002)}]{2002P&SS...50.1197G}
{Galli}, D. \& {Palla}, F. 2002, \planss, 50, 1197

\bibitem[{{Glover}(2005)}]{2005SSRv..117..445G}
{Glover}, S. 2005, Space Science Reviews, 117, 445

\bibitem[{{Glover} \& {Abel}(2008)}]{2008MNRAS.388.1627G}
{Glover}, S.~C.~O. \& {Abel}, T. 2008, \mnras, 388, 1627

\bibitem[{{Glover} {et~al.}(2008){Glover}, {Clark}, {Greif}, {Johnson},
  {Bromm}, {Klessen}, \& {Stacy}}]{2008IAUS..255....3G}
{Glover}, S.~C.~O., {Clark}, P.~C., {Greif}, T.~H., {et~al.} 2008, in IAU
  Symposium, ed. {L.~K.~Hunt, S.~Madden, \& R.~Schneider}, Vol. 255, 3--17

\bibitem[{{Gosachinskij} {et~al.}(2002){Gosachinskij}, {Dubrovich},
  {Zhelenkov}, {Il'in}, \& {Prozorov}}]{2002ARep...46..543G}
{Gosachinskij}, I.~V., {Dubrovich}, V.~K., {Zhelenkov}, S.~R., {Il'in}, G.~N.,
  \& {Prozorov}, V.~A. 2002, Astronomy Reports, 46, 543

\bibitem[{{Greif} {et~al.}(2008{\natexlab{a}}){Greif}, {Johnson}, \&
  {Bromm}}]{2008AIPC..990..405G}
{Greif}, T.~H., {Johnson}, J.~L., \& {Bromm}, V. 2008{\natexlab{a}}, in
  American Institute of Physics Conference Series, Vol. 990, First Stars III,
  ed. {B.~W.~O'Shea \& A.~Heger}, 405--417

\bibitem[{{Greif} {et~al.}(2008{\natexlab{b}}){Greif}, {Johnson}, {Klessen}, \&
  {Bromm}}]{2008MNRAS.387.1021G}
{Greif}, T.~H., {Johnson}, J.~L., {Klessen}, R.~S., \& {Bromm}, V.
  2008{\natexlab{b}}, \mnras, 387, 1021

\bibitem[{{Haiman} {et~al.}(2000){Haiman}, {Abel}, \&
  {Rees}}]{2000ApJ...534...11H}
{Haiman}, Z., {Abel}, T., \& {Rees}, M.~J. 2000, \apj, 534, 11

\bibitem[{{Haiman} {et~al.}(1997){Haiman}, {Rees}, \&
  {Loeb}}]{1997ApJ...476..458H}
{Haiman}, Z., {Rees}, M.~J., \& {Loeb}, A. 1997, \apj, 476, 458

\bibitem[{{Hogan} \& {Rees}(1979)}]{1979MNRAS.188..791H}
{Hogan}, C.~J. \& {Rees}, M.~J. 1979, \mnras, 188, 791

\bibitem[{{Jee} {et~al.}(2009){Jee}, {Rosati}, {Ford}, {Dawson}, {Lidman},
  {Perlmutter}, {Demarco}, {Strazzullo}, {Mullis}, {B{\"o}hringer}, \&
  {Fassbender}}]{2009ApJ...704..672J}
{Jee}, M.~J., {Rosati}, P., {Ford}, H.~C., {et~al.} 2009, \apj, 704, 672

\bibitem[{{Jenkins} {et~al.}(1998){Jenkins}, {Frenk}, {Pearce}, {Thomas},
  {Colberg}, {White}, {Couchman}, {Peacock}, {Efstathiou}, \&
  {Nelson}}]{1998ApJ...499...20J}
{Jenkins}, A., {Frenk}, C.~S., {Pearce}, F.~R., {et~al.} 1998, \apj, 499, 20

\bibitem[{{Johnson} \& {Bromm}(2006)}]{2006MNRAS.366..247J}
{Johnson}, J.~L. \& {Bromm}, V. 2006, \mnras, 366, 247

\bibitem[{{Kamaya} \& {Silk}(2003)}]{2003MNRAS.339.1256K}
{Kamaya}, H. \& {Silk}, J. 2003, \mnras, 339, 1256

\bibitem[{{Komatsu} {et~al.}(2009){Komatsu}, {Dunkley}, {Nolta}, {Bennett},
  {Gold}, {Hinshaw}, {Jarosik}, {Larson}, {Limon}, {Page}, {Spergel},
  {Halpern}, {Hill}, {Kogut}, {Meyer}, {Tucker}, {Weiland}, {Wollack}, \&
  {Wright}}]{2009ApJS..180..330K}
{Komatsu}, E., {Dunkley}, J., {Nolta}, M.~R., {et~al.} 2009, \apjs, 180, 330

\bibitem[{{Kurki-Suonio} {et~al.}(1997){Kurki-Suonio}, {Jedamzik}, \&
  {Mathews}}]{1997ApJ...479...31K}
{Kurki-Suonio}, H., {Jedamzik}, K., \& {Mathews}, G.~J. 1997, \apj, 479, 31

\bibitem[{{Lara} {et~al.}(2006){Lara}, {Kajino}, \&
  {Mathews}}]{2006PhRvD..73h3501L}
{Lara}, J.~F., {Kajino}, T., \& {Mathews}, G.~J. 2006, \prd, 73, 083501

\bibitem[{{Lepp} \& {Shull}(1984)}]{1984ApJ...280..465L}
{Lepp}, S. \& {Shull}, J.~M. 1984, \apj, 280, 465

\bibitem[{{Lepp} {et~al.}(2002){Lepp}, {Stancil}, \&
  {Dalgarno}}]{2002JPhB...35R..57L}
{Lepp}, S., {Stancil}, P.~C., \& {Dalgarno}, A. 2002, Journal of Physics B
  Atomic Molecular Physics, 35, 57

\bibitem[{{Loeb}(2008)}]{2008arXiv0804.2258L}
{Loeb}, A. 2008, arXiv: 0804.2258

\bibitem[{{Longair}(2008)}]{2008gafo.book.....L}
{Longair}, M.~S. 2008, {Galaxy Formation} (Galaxy Formation, by Malcolm
  S.~Longair Berlin: Springer, 2008.~ ISBN 978-3-540-73477-2)

\bibitem[{{Mackey} {et~al.}(2003){Mackey}, {Bromm}, \&
  {Hernquist}}]{2003ApJ...586....1M}
{Mackey}, J., {Bromm}, V., \& {Hernquist}, L. 2003, \apj, 586, 1

\bibitem[{{Maoli} {et~al.}(1996){Maoli}, {Ferrucci}, {Melchiorri}, {Signore},
  \& {Tosti}}]{1996ApJ...457....1M}
{Maoli}, R., {Ferrucci}, V., {Melchiorri}, F., {Signore}, M., \& {Tosti}, D.
  1996, \apj, 457, 1

\bibitem[{{Maoli} {et~al.}(1994){Maoli}, {Melchiorri}, \&
  {Tosti}}]{1994ApJ...425..372M}
{Maoli}, R., {Melchiorri}, F., \& {Tosti}, D. 1994, \apj, 425, 372

\bibitem[{{Mizusawa} {et~al.}(2004){Mizusawa}, {Nishi}, \&
  {Omukai}}]{2004PASJ...56..487M}
{Mizusawa}, H., {Nishi}, R., \& {Omukai}, K. 2004, \pasj, 56, 487

\bibitem[{{Nordh} {et~al.}(2003){Nordh}, {von Sch{\'e}ele}, {Frisk}, {Ahola},
  {Booth}, {Encrenaz}, {Hjalmarson}, {Kendall}, {Kyr{\"o}l{\"a}}, {Kwok},
  {Lecacheux}, {Leppelmeier}, {Llewellyn}, {Mattila}, {M{\'e}gie}, {Murtagh},
  {Rougeron}, \& {Witt}}]{2003A&A.402.21.Nordh.etal}
{Nordh}, H.~L., {von Sch{\'e}ele}, F., {Frisk}, U., {et~al.} 2003, \aap, 402,
  L21

\bibitem[{{Okamoto}(2008)}]{2008ASPC..393..111O}
{Okamoto}, T. 2008, in Astronomical Society of the Pacific Conference Series,
  Vol. 393, New Horizons in Astronomy, ed. {A.~Frebel, J.~R.~Maund, J.~Shen, \&
  M.~H.~Siegel}, 111

\bibitem[{{Olberg} {et~al.}(2003){Olberg}, {Frisk}, {Lecacheux}, {Olofsson},
  {Baron}, {Bergman}, {Florin}, {Hjalmarson}, {Larsson}, {Murtagh}, {Olofsson},
  {Pagani}, {Sandqvist}, {Teyssier}, {Torchinsky}, \&
  {Volk}}]{2003A&A.402.35O.Olberg.etal}
{Olberg}, M., {Frisk}, U., {Lecacheux}, A., {et~al.} 2003, \aap, 402, L35

\bibitem[{{Olofsson} {et~al.}(2007){Olofsson}, {Persson}, {Koning}, {Bergman},
  {Bernath}, {Black}, {Frisk}, {Geppert}, {Hasegawa}, {Hjalmarson}, {Kwok},
  {Larsson}, {Lecacheux}, {Nummelin}, {Olberg}, {Sandqvist}, \&
  {Wirstr{\"o}m}}]{2007A&A...476..791O}
{Olofsson}, A.~O.~H., {Persson}, C.~M., {Koning}, N., {et~al.} 2007, \aap, 476,
  791

\bibitem[{{Omukai} \& {Kitayama}(2003)}]{2003ApJ...599..738O}
{Omukai}, K. \& {Kitayama}, T. 2003, \apj, 599, 738

\bibitem[{{Papadopoulos} {et~al.}(2001){Papadopoulos}, {Ivison}, {Carilli}, \&
  {Lewis}}]{2001Natur.409...58P}
{Papadopoulos}, P., {Ivison}, R., {Carilli}, C., \& {Lewis}, G. 2001, \nat,
  409, 58

\bibitem[{{Persson} {et~al.}(2009){Persson}, {Olberg}, {.~Hjalmarson},
  {Spaans}, {Black}, {Frisk}, {Liljestr{\"o}m}, {Olofsson}, {Poelman}, \&
  {Sandqvist}}]{2009A&A...494..637P}
{Persson}, C.~M., {Olberg}, M., {.~Hjalmarson}, {\AA}., {et~al.} 2009, \aap,
  494, 637

\bibitem[{{Persson} {et~al.}(2007){Persson}, {Olofsson}, {Koning}, {Bergman},
  {Bernath}, {Black}, {Frisk}, {Geppert}, {Hasegawa}, {Hjalmarson}, {Kwok},
  {Larsson}, {Lecacheux}, {Nummelin}, {Olberg}, {Sandqvist}, \&
  {Wirstr{\"o}m}}]{2007A&A...476..807PaperII}
{Persson}, C.~M., {Olofsson}, A.~O.~H., {Koning}, N., {et~al.} 2007, \aap, 476,
  807

\bibitem[{{Prochaska} {et~al.}(2009){Prochaska}, {Sheffer}, {Perley}, {Bloom},
  {Lopez}, {Dessauges-Zavadsky}, {Chen}, {Filippenko}, {Ganeshalingam}, {Li},
  {Miller}, \& {Starr}}]{2009ApJ...691L..27P}
{Prochaska}, J.~X., {Sheffer}, Y., {Perley}, D.~A., {et~al.} 2009, \apjl, 691,
  L27

\bibitem[{{Puy} {et~al.}(1993){Puy}, {Alecian}, {Le Bourlot}, {Leorat}, \&
  {Pineau Des Forets}}]{1993A&A...267..337P}
{Puy}, D., {Alecian}, G., {Le Bourlot}, J., {Leorat}, J., \& {Pineau Des
  Forets}, G. 1993, \aap, 267, 337

\bibitem[{{Puy} {et~al.}(2007){Puy}, {Dubrovich}, {Lipovka}, {Talbi}, \&
  {Vonlanthen}}]{2007A&A...476..685P}
{Puy}, D., {Dubrovich}, V., {Lipovka}, A., {Talbi}, D., \& {Vonlanthen}, P.
  2007, \aap, 476, 685

\bibitem[{{Puy} \& {Signore}(1996)}]{1996A&A...305..371P}
{Puy}, D. \& {Signore}, M. 1996, \aap, 305, 371

\bibitem[{{Puy} \& {Signore}(2007)}]{2007NewAR..51..411P}
{Puy}, D. \& {Signore}, M. 2007, New Astronomy Review, 51, 411

\bibitem[{{Rauscher} {et~al.}(1994){Rauscher}, {Applegate}, {Cowan},
  {Thielemann}, \& {Wiescher}}]{1994ApJ...429..499R}
{Rauscher}, T., {Applegate}, J.~H., {Cowan}, J.~J., {Thielemann}, F.-K., \&
  {Wiescher}, M. 1994, \apj, 429, 499

\bibitem[{{Rohlfs} \& {Wilson}(2004)}]{2004tra..book.....R}
{Rohlfs}, K. \& {Wilson}, T.~L. 2004, {Tools of radio astronomy}, ed. {Rohlfs,
  K.~\& Wilson, T.~L.}

\bibitem[{{Rubi{\~n}o-Mart{\'{\i}}n} {et~al.}(2006){Rubi{\~n}o-Mart{\'{\i}}n},
  {Chluba}, \& {Sunyaev}}]{2006MNRAS.371.1939R}
{Rubi{\~n}o-Mart{\'{\i}}n}, J.~A., {Chluba}, J., \& {Sunyaev}, R.~A. 2006,
  \mnras, 371, 1939

\bibitem[{{Rubi{\~n}o-Mart{\'{\i}}n} {et~al.}(2008){Rubi{\~n}o-Mart{\'{\i}}n},
  {Chluba}, \& {Sunyaev}}]{2008A&A...485..377R}
{Rubi{\~n}o-Mart{\'{\i}}n}, J.~A., {Chluba}, J., \& {Sunyaev}, R.~A. 2008,
  \aap, 485, 377

\bibitem[{{Santoro} \& {Shull}(2006)}]{2006ApJ...643...26S}
{Santoro}, F. \& {Shull}, J.~M. 2006, \apj, 643, 26

\bibitem[{{Schleicher} {et~al.}(2008){Schleicher}, {Galli}, {Palla},
  {Camenzind}, {Klessen}, {Bartelmann}, \& {Glover}}]{2008A&A...490..521S}
{Schleicher}, D.~R.~G., {Galli}, D., {Palla}, F., {et~al.} 2008, \aap, 490, 521

\bibitem[{{Schneider} {et~al.}(2004){Schneider}, {Ferrara}, \&
  {Salvaterra}}]{2004MNRAS.351.1379S}
{Schneider}, R., {Ferrara}, A., \& {Salvaterra}, R. 2004, \mnras, 351, 1379

\bibitem[{{Sethi} {et~al.}(2008){Sethi}, {Nath}, \&
  {Subramanian}}]{2008MNRAS.387.1589S}
{Sethi}, S.~K., {Nath}, B.~B., \& {Subramanian}, K. 2008, \mnras, 387, 1589

\bibitem[{{Sethi} {et~al.}(2007){Sethi}, {Subrahmanyan}, \&
  {Roshi}}]{2007ApJ...664....1S}
{Sethi}, S.~K., {Subrahmanyan}, R., \& {Roshi}, D.~A. 2007, \apj, 664, 1

\bibitem[{{Signore} \& {Puy}(2009)}]{2009EPJC...59..117S}
{Signore}, M. \& {Puy}, D. 2009, European Physical Journal C, 59, 117

\bibitem[{{Silk}(1968)}]{1968ApJ...151..459S}
{Silk}, J. 1968, \apj, 151, 459

\bibitem[{{Smith} {et~al.}(2009){Smith}, {Turk}, {Sigurdsson}, {O'Shea}, \&
  {Norman}}]{2009ApJ...691..441S}
{Smith}, B.~D., {Turk}, M.~J., {Sigurdsson}, S., {O'Shea}, B.~W., \& {Norman},
  M.~L. 2009, \apj, 691, 441

\bibitem[{{Srianand} {et~al.}(2008){Srianand}, {Noterdaeme}, {Ledoux}, \&
  {Petitjean}}]{2008A&A...482L..39S}
{Srianand}, R., {Noterdaeme}, P., {Ledoux}, C., \& {Petitjean}, P. 2008, \aap,
  482, L39

\bibitem[{{Stanway} {et~al.}(2008){Stanway}, {Bremer}, {Davies}, {Birkinshaw},
  {Douglas}, \& {Lehnert}}]{2008ApJ...687L...1S}
{Stanway}, E.~R., {Bremer}, M.~N., {Davies}, L.~J.~M., {et~al.} 2008, \apjl,
  687, L1

\bibitem[{{Steigman}(2007)}]{2007ARNPS..57..463S}
{Steigman}, G. 2007, Annual Review of Nuclear and Particle Science, 57, 463

\bibitem[{{Tanvir} {et~al.}(2009){Tanvir}, {Fox}, {Levan}, {Berger},
  {Wiersema}, {Fynbo}, {Cucchiara}, {Kr{\"u}hler}, {Gehrels}, {Bloom},
  {Greiner}, {Evans}, {Rol}, {Olivares}, {Hjorth}, {Jakobsson}, {Farihi},
  {Willingale}, {Starling}, {Cenko}, {Perley}, {Maund}, {Duke}, {Wijers},
  {Adamson}, {Allan}, {Bremer}, {Burrows}, {Castro-Tirado}, {Cavanagh}, {de
  Ugarte Postigo}, {Dopita}, {Fatkhullin}, {Fruchter}, {Foley}, {Gorosabel},
  {Kennea}, {Kerr}, {Klose}, {Krimm}, {Komarova}, {Kulkarni}, {Moskvitin},
  {Mundell}, {Naylor}, {Page}, {Penprase}, {Perri}, {Podsiadlowski}, {Roth},
  {Rutledge}, {Sakamoto}, {Schady}, {Schmidt}, {Soderberg}, {Sollerman},
  {Stephens}, {Stratta}, {Ukwatta}, {Watson}, {Westra}, {Wold}, \&
  {Wolf}}]{2009Natur.461.1254T}
{Tanvir}, N.~R., {Fox}, D.~B., {Levan}, A.~J., {et~al.} 2009, \nat, 461, 1254

\bibitem[{{Vonlanthen} \& {Puy}(2008)}]{2008sf2a.conf..355V}
{Vonlanthen}, P. \& {Puy}, D. 2008, in SF2A-2008: Proceedings of the Annual
  meeting of the French Society of Astronomy and Astrophysics, ed.
  C.~{Charbonnel}, F.~{Combes}, \& R.~{Samadi}, 355

\bibitem[{{Vonlanthen} {et~al.}(2009){Vonlanthen}, {Rauscher}, {Winteler},
  {Puy}, {Signore}, \& {Dubrovich}}]{2009A&A...503...47V}
{Vonlanthen}, P., {Rauscher}, T., {Winteler}, C., {et~al.} 2009, \aap, 503, 47

\bibitem[{{Wise} \& {Abel}(2008)}]{2008ApJ...685...40W}
{Wise}, J.~H. \& {Abel}, T. 2008, \apj, 685, 40

\bibitem[{{Woosley} \& {Bloom}(2006)}]{2006ARA&A..44..507W}
{Woosley}, S.~E. \& {Bloom}, J.~S. 2006, \araa, 44, 507

\end{thebibliography}
 \Online
  \appendix
  \section{Figures}
  
\begin{figure} [\!ht]
\centering
\includegraphics[scale=0.5]{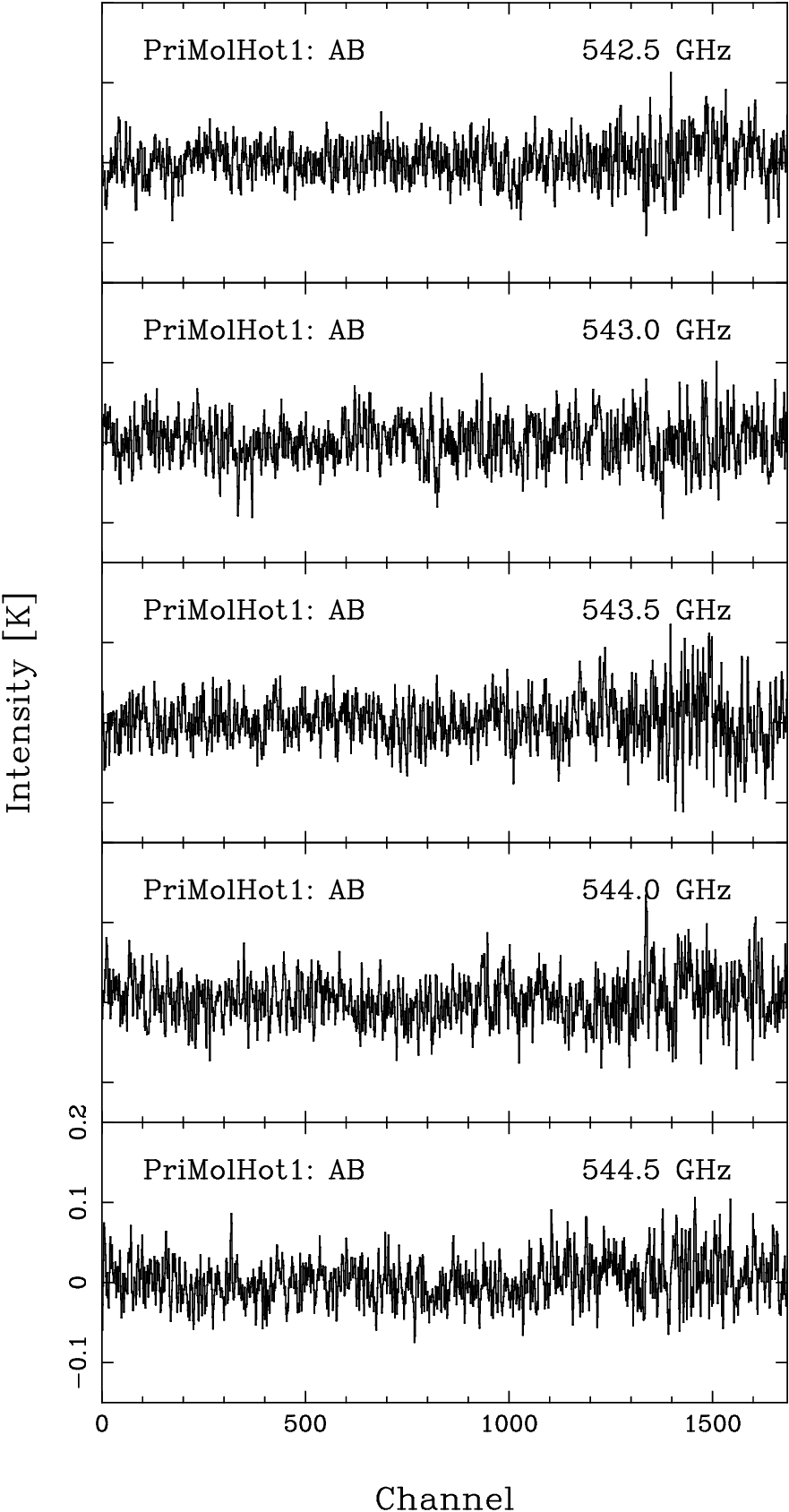}
 \caption{Spectra from weekend 1\,--\,5 vs. channels for Position A with B as reference from the 2006/07 survey in original shape and channel spacing  0.62\,MHz.}
 \label{Fig: AB week 1-5 vs. channel}
\end{figure}

\begin{figure}[\!ht] 
\centering
\includegraphics[scale=0.5]{13395Fg11.eps}  
 \caption{Spectra from weekend 6\,--\,10 vs. channels for Position A with B as reference from the 2006/07 survey in original shape and channel spacing  0.62\,MHz.}
 \label{Fig: AB week 6-10 vs. channel}
\end{figure}

\begin{figure} 
\includegraphics[scale=0.5]{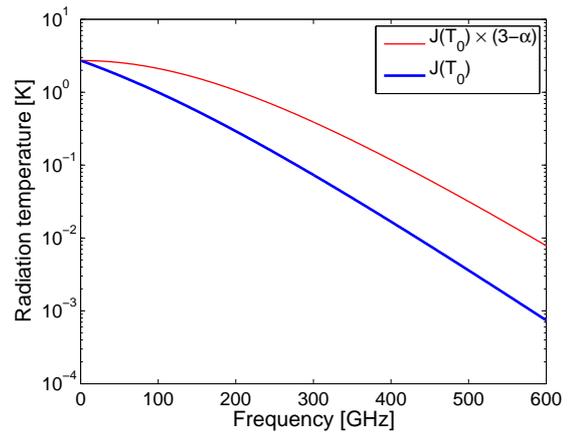}  
\caption{The blue thick line shows the radiation temperature $J(T_0)$, and the red thin line shows $J(T_0)\times(3-\alpha$) as a function of frequency for  $T_\mathrm{0}$\,=\,2.725\,K.
}
 \label{Fig: Jx(3-alpha) vs freq}
\end{figure}

\clearpage
\section{The turn-around redshift as a function of density fluctuations on
the last scattering surface} \label{appendix: derivation TA(z)}

Let the unperturbed   density at the last scattering surface (LSS) be
$\rho_{\mathrm{LSS}} = \rho_\mathrm{m_\mathrm{\,LSS}} + \rho_{\Lambda_\mathrm{\,LSS}}
+\rho_{\mathrm{rad}_\mathrm{\,LSS}}$, where   $\rho_\mathrm{m_{\,LSS}}$ is the matter density, $ \rho_\mathrm{\Lambda_\mathrm{\,LSS}}$ is the dark energy density and $\rho_\mathrm{rad_{\,LSS}}$ the
radiation energy density. Since the matter
density is very dominant at LSS $\rho_{\mathrm{LSS}}  \approx \rho_\mathrm{m_{\,LSS}}$.
This implies that the 
critical density 
$  \rho_{\mathrm{crit}_\mathrm{\,LSS}}=  \rho_\mathrm{m_{\,LSS}} + \rho_\mathrm{\,\Lambda_{\,LSS}}
+\rho_\mathrm{rad_{\,LSS}} + \rho_{k_\mathrm{\,LSS}} \approx \rho_\mathrm{m_{\,LSS}}$  since we assume that the curvature density   $\rho_\mathrm{k_{\,LSS}}$ is zero. 
The   density parameter at the LSS  then becomes $\Omega_\mathrm{{LSS}}= \rho_\mathrm{LSS} /\rho_\mathrm{crit_\mathrm{\,LSS}} = 1$.

Let a perturbed density at LSS  $\rho^\prime_\mathrm{{LSS}}$ with the (large)  scale $a_\mathrm{LSS}$
be 
defined by
\begin{equation}
\rho^\prime_\mathrm{{LSS}}=\rho_\mathrm{LSS}+\Delta \rho_\mathrm{LSS}\ ,
\end{equation}
where
$\Delta \rho_\mathrm{LSS}$ is the   density perturbation.
If $\Delta \rho_\mathrm{LSS}$ is positive, the turn-around scale $a^\prime_\mathrm{\,TA}$ of the
perturbed region is  \citep[cf.][Eq.~(16.2)]{2008gafo.book.....L}
\begin{equation}
a^\prime_\mathrm{TA} =a_\mathrm{LSS} \frac{\Omega^\prime_\mathrm{{LSS}}} {\Omega^\prime_\mathrm{{LSS}}-1}\ , 
\end{equation}
where
 \begin{equation}
 \Omega^\prime_\mathrm{{LSS}}=\frac{\rho_\mathrm{LSS}+\Delta \rho_\mathrm{LSS}}{\rho_\mathrm{LSS}}=
 1+\frac{\Delta \rho_\mathrm{LSS}}{\rho_\mathrm{LSS}}\ .
 \end{equation}
If $\Delta \rho_\mathrm{LSS}$ is negative, there is of course no turn-around.

The scale $a_\mathrm{TA}$ in the unperturbed region at the turn-around time $t=t_\mathrm{TA}$ 
is then (cf. Longair \citep[cf.][Eq.~(16.3)]{2008gafo.book.....L}
\begin{equation}
a_{TA}=5.55^{1/3} \cdot a^\prime_\mathrm{TA}=5.55^{1/3}\cdot a_\mathrm{LSS} \frac{\Omega^\prime_\mathrm{LSS}}
 {\Omega^\prime_\mathrm{LSS}-1}\ .
 \end{equation}

  Since $a_\mathrm{TA}=(1+z_\mathrm{TA})^{-1}$ and $a_\mathrm{LSS}=(1+z_\mathrm{LSS})^{-1}$ we have
 \begin{equation}
(1+z_\mathrm{TA})=(1+z_\mathrm{LSS}) \cdot 5.55^{-1/3} \frac{\Delta \rho_\mathrm{LSS}}{\rho_\mathrm{LSS}+\Delta
\rho_\mathrm{LSS}}\ .
 \end{equation}
 If $\Delta \rho_\mathrm{LSS} \ll \rho_\mathrm{LSS}$ we have
 \begin{equation}
 (1+z_\mathrm{TA})=(1+z_\mathrm{LSS}) \cdot 5.55^{-1/3} \frac{\Delta \rho_\mathrm{LSS}}{\rho_\mathrm{LSS}}\ .
 \end{equation}

We may replace $\Delta \rho_\mathrm{LSS} / \rho_\mathrm{LSS}$ with 
$\Delta M_\mathrm{LSS}/M_\mathrm{LSS}$, where $M$ is the mass within the scale $a$, 
 so that
\begin{equation}
 \label{Zturn}
 (1+z_\mathrm{TA})=(1+z_\mathrm{LSS}) \cdot 5.55^{-1/3} \frac{\Delta M_\mathrm{LSS}}{M_\mathrm{LSS}}\ .
\end{equation}
 Consider now density fluctuations of the Harrison-Zeldovich type \citep[cf.][page 390\,--\,392]{2008gafo.book.....L}, where
 \begin{equation}
 \label{sigma}
 \sigma_{M}=\left < \left ( \frac{\Delta M}{M}\right )^{2} \right>^{1/2} \sim M^{-2/3}\ .
 \end{equation}
 Note that this relation may be expressed as
 \begin{equation}
 \label{BH}
  < (\Delta M) ^{2}>^{1/2} \sim M^{1/3}\ .
 \end{equation}
assuring that black holes are not formed excessively on small or large scales.

 Assume now that Eq.~(\ref{sigma}) can be normalised by the observed fluctuations
$\sigma_{M_{S}}=10^{-4}$ of the mass \mbox{$M_{S}=3.72 \times 10^{15}\,M_{\odot}$}  \citep[cf.][Eq.~(15.13)]{2008gafo.book.....L} 
 within the sound horizon at the last scattering surface where $z_\mathrm{LSS}=1\,090$.
Equation
 (\ref{sigma}) is then normalised to
 \begin{equation}
 \label{sigma2}
 \sigma_{M}=10^{-4} \left ( \frac {M_{S}}{M} \right )^{2/3}\ .
 \end{equation}

Expressing  $\Delta M/M$ in Eq.~(\ref{Zturn}) in terms of $\xi \cdot \sigma_{M}$, where $\xi$ is a sigma measure, we
get
\begin{equation}
 (1+z_\mathrm{TA})=(1+z_\mathrm{LSS}) \cdot 5.55^{-1/3} \cdot \xi  \cdot  \sigma_{M}\ .
\end{equation}  
 Note that $\xi>0$.

 Inserting Eq.~(\ref{sigma2}) and $M_{S}$ we get,
 \begin{equation}\label{final turn-around z}
 1+z_\mathrm{TA}=(1+z_\mathrm{LSS}) \cdot 1.35 \cdot \xi  \cdot \left ( \frac{10^{9}\, 
\mathrm{M}_{\odot}}{M}\right )^{2/3}\ .
 \end{equation}

This derivation is valid for dark matter. Baryon structure formation, also driven by gravitational forces, has 
as well been affected by Silk damping \citep{1968ApJ...151..459S} and sound waves. This complication introduces a minor modification to the 
power spectrum of the density fluctuations \citep[page 412 in][]{2008gafo.book.....L}, but approximatively our derivation should hold also for 
the baryonic matter. More worrisome for the Harrison-Zeldovich scenario are perhaps the   findings of  
massive galaxy clusters at high redshifts, for instance XMMU J2235.3-2557  at $z\!=\!1.4$ with a estimated projected mass of the cluster within  1\,Mpc  of  $8.5\pm1.7\times10^{14}$\,M$_\odot$~\citep{2009ApJ...704..672J} which according to Eq.~(\ref{final turn-around z}) constitutes a 
14$\sigma$ event.

It should perhaps be noted that masses on the order of $10^8-10^9$\,M$_\odot$~have turnaround redshifts on the order of  $10^3-10^4$. This means that black holes of this size could form shortly thereafter if the collapse is not halted by induced angular momentum.


\section{Tables}

\begin{table*} [!ht]
\caption{The ranges in redshifts covered by our observations for the five lowest
rotational lines of   H$_2$,  HD, and HeH$^+$.}
\begin{tabular} {lrcrccc   cc}
\hline
\hline
   \noalign{\smallskip}
Species&  Frequency & Transitions& 	E$_u$& A-coeff	 &AC [GHz]&  AOS [GHz] &  AOS [GHz] & AC	 [GHz]\\
 \noalign{\smallskip}
& [GHz]  & $J_u$\,--\,$J_l$&[K] & [s$^{-1}$]  & 486.5\,--\,492.0& 542.0\,--\,547.5&  547.0\,--\,563.0 &  563.0\,--\,578.0\\
  \noalign{\smallskip}
\hline
   \noalign{\smallskip}
 
H$_2$&10\,621    & 2\,--\,0	&510  &2.94\x10$^{-11}$&	$z$\,=\,20.83\,--\,20.59	& 18.60\,--\,18.40	&	18.42\,--\,17.87	&	17.87\,--\,17.38 \\
&17\,594  &3\,--\,1	&1\,015&4.76\x10$^{-10}$ &	$z$\,=\,35.16\,--\,34.76& 31.46\,--\,31.14	&	31.16\,--\,30.25	&	30.25\,--\,29.44 \\
&24\,410 &4\,--\,2	&1\,681&2.75\x10$^{-9}$&	$z$\,=\,49.18\,--\,48.61	& 44.04\,--\,43.58	&	43.63\,--\,42.36	&	42.36\,--\,41.23 \\
&31\,011 & 5\,--\,3	&2\,503 &9.83\x10$^{-9}$ &	$z$\,=\,62.74\,--\,62.03	& 56.22\,--\,55.64	&	55.69\,--\,54.08	&	54.08\,--\,52.65\\
&37\,348  &6\,--\,4	&3\,474&2.64\x10$^{-8}$&	$z$\,=\,75.77\,--\,74.91	& 67.91\,--\,67.22	&	67.28\,--\,65.34	&	65.34\,--\,63.62 \\

   \noalign{\smallskip}
HD &2\,675 & 1\,--\,0	&128&5.32\x10$^{-8}$&	$z$\,=\,4.50\,--\,4.37	& 3.94\,--\,3.89	&	3.89\,--\,3.75	&	3.75\,--\,3.63 \\
&5\,332 & 2\,--\,1	&384&5.05\x10$^{-7}$&	$z$\,=\,9.96\,--\,9.84	& 8.84\,--\,8.74	&	8.75\,--\,8.47	&	8.47\,--\,8.22 \\
&7\,952 & 3\,--\,2	&766&1.80\x10$^{-6}$&	$z$\,=\,15.35\,--\,15.16	& 13.67\,--\,13.52	&	13.54\,--\,13.12	&	13.12\,--\,12.76 \\
&10\,518  &4\,--\,3	&1\,271&4.31\x10$^{-6}$&	$z$\,=\,20.62\,--\,20.38	& 18.41\,--\,18.21	&	18.23\,--\,17.68	&	17.68\,--\,17.20 \\
&13\,015  &5\,--\,4	&1\,895&8.35\x10$^{-6}$&	$z$\,=\,25.75\,--\,25.45	& 23.01\,--\,22.77	&	22.79\,--\,22.12	&	22.12\,--\,21.52 \\

  \noalign{\smallskip} 
  
  HeH$^+$ &2\,010 &  1\,--\,0	&96 &0.109 &	$z$\,=\,3.13\,--\,3.09	& 2.71\,--\,2.67	&	2.67\,--\,2.57	&	2.57\,--\,2.48 \\
&4\,009  &2\,--\,1	&289&1.04&	$z$\,=\,7.24\,--\,7.15	& 6.40\,--\,6.32	&	6.33\,--\,6.12	&	6.12\,--\,5.94 \\
&5\,984 & 3\,--\,2	&576&3.75&	$z$\,=\,11.30\,--\,11.16	& 10.04\,--\,9.93	&	9.94\,--\,9.63	&	9.63\,--\,9.35 \\
&7\,925 & 4\,--\,3	&956&9.14&	$z$\,=\,15.29\,--\,15.11	& 13.62\,--\,13.48	&	13.49\,--\,13.08	&	13.08\,--\,12.71 \\
&9\,821 &5\,--\,4	&1\,428&18.1&	$z$\,=\,19.19\,--\,18.96	& 17.12\,--\,16.94	&	16.95\,--\,16.44	&	16.44\,--\,15.99 \\
 \noalign{\smallskip}
  \hline
\end{tabular}
\label{Table: redshift coverage species} 
\end{table*}

\begin{table} [!hb]
\caption{Physical sizes   corresponding  to the Odin beam and the different angular steps in the 2006/07 observations.}
\centering
\begin{tabular}{r r rrrrrr}
 \noalign{\smallskip}
\hline
\hline
 \noalign{\smallskip}
Angular    &\multicolumn{7}{c}{Redshift $z$}\\
 size  & $z$\,=\,5  &  10&25 & 50& 100   & 200  &300\\
\hline
\noalign{\smallskip}
$\theta$ [\arcmin]  & \multicolumn{7}{c}{Physical size [kpc]}\\
\noalign{\smallskip}
\hline
\noalign{\smallskip}
 2$\farcm$1 &820 & 	540&270& 150&79&41& 28\\
5& 2\,000  &		1\,300&640& 350 & 190 &  98 &   66 \\
10 & 3\,900  & 		2\,600&1\,300& 700 & 380 &  200  &  130\\
15 & 5\,900  & 		3\,900&1\,900& 1\,060 & 560 &  290 &  200 \\
30 &	12\,000 &	7\,700& 3\,800& 2\,100 & 1\,100   & 590 &   400\\
\noalign{\smallskip}
\hline
\end{tabular}
\label{Odin step scale} 
\end{table}

\begin{table*} [!ht]
\caption{Line widths in the linear   phase for  a perturbation corresponding to the Odin beam at a number of redshifts.}
\centering
\begin{tabular}{ccccccc  }
 \noalign{\smallskip}
\hline
\hline
 \noalign{\smallskip}
Redshift	&	Temp	& Size$^{{a}}$	& Mass$^{{a}}$& 	Fract. width 	&   Line width$^b$	&	Width at 500\,GHz \\
\noalign{\smallskip}
$z$	& $T$	&	$L$	&	$M$  	&$\Delta \nu/\nu$& $\Delta \upsilon_\mathrm{Linear}$& $\Delta \nu$  	\\
\noalign{\smallskip}
&[K]	&	[kpc]	&	[M$_\odot$] 	&& [\kms]	&[GHz]	 ]\\
\noalign{\smallskip}
\hline
\noalign{\smallskip}

10	& 30  &	540 &	4\x10$^{12}$ &	 		2\x10$^{-3}$ & 720	& 1.3	 \\
20	& 57  &  320 &	6\x10$^{12}$ &	 	4\x10$^{-3}$ & 1\,100	& 2.1	  \\
50	& 139  &  150 &	9\x10$^{12}$ &	 	7\x10$^{-3}$ &2\,000	 & 3.7	 	\\
75	& 207  &  100 &	9\x10$^{12}$ &	 	8\x10$^{-3}$ &2\,500	& 4.5	 	\\
100	& 275  &  79 &		1\x10$^{13}$ &	 	1\x10$^{-2}$ &2\,900	& 5.3	 	\\
150	& 411  &  54 &		1\x10$^{13}$ &	 	1\x10$^{-2}$ &3\,600	& 6.6	 		\\
200	& 547  &  41 &		1\x10$^{13}$ &	 	1\x10$^{-2}$ &4\,200	& 7.7	 	\\

\noalign{\smallskip}
\hline
\end{tabular}
\begin{list}{}{}
\item$^{{a}}$For an angular size of the Odin beam, 2$\farcm$1 calculated with Eqs.~(\ref{angular size})\,--\,(\ref{mass within perturbation}). $^b$The line width during the linear evolution, Eq.~(\ref{fraction of resolution}).
\end{list}
\label{Table: line widths} 
\end{table*}

 \end{document}